\documentclass[aps,
pra,
twocolumn,
superscriptaddress,
amssymb,
notitlepage,
reprint,
10pt]{revtex4-1}

\usepackage{upgreek}

\usepackage{amsmath}

\usepackage{graphicx}

\newcommand{\mode}[1]{\hat{#1}}

\newcommand{\modea}{\mode{a}}

\newcommand{\be}{\begin{equation}}
\newcommand{\ee}{\end{equation}}

\newcommand{\ket}[1]{|#1\rangle}
\newcommand{\ketL}[1]{|#1\rangle_L}

\newcommand{\ko}{\kappa_\text{out}}

\newcommand{\bo}{b_\text{out}}
\newcommand{\bi}{b_\text{in}}
\newcommand{\chiab}{\chi_\text{ab}}
\newcommand{\chiac}{\chi_\text{at}}
\newcommand{\chibc}{\chi_\text{bt}}
\newcommand{\chiaa}{\chi_\text{aa}}
\newcommand{\chibb}{\chi_\text{bb}}
\newcommand{\chicc}{\chi_\text{tt}}
\newcommand{\deltaa}{\delta_\text{a}}
\newcommand{\deltab}{\delta_\text{b}}

\newcommand{\eqnref}{Equation\,\,}

\newcommand{\figref}{Figure\,\,}
\newcommand{\tabref}{Table\,\,}

\graphicspath{{./figures/}{./supplement_figures/}}

\newenvironment{sciabstract}{%
\begin{abstract}}
{\end{abstract}}

\newcommand{\supp}{(see Supplement)} 

\begin{document}
\begin{sciabstract}
  Modular quantum computing architectures require fast and efficient distribution of quantum information through propagating signals. 
  Here we report rapid, on-demand quantum state transfer between two remote superconducting cavity quantum memories through traveling microwave photons.
  We demonstrate a quantum communication channel by deterministic transfer of quantum bits with 76\% fidelity. 
  Heralding on errors induced by experimental imperfection can improve this to 87\% with a success probability of 0.87.
  By partial transfer of a microwave photon, we generate remote entanglement at a rate that exceeds photon loss in either memory by more than a factor of three.
  We further show the transfer of quantum error correction code words that will allow deterministic mitigation of photon loss.
  These results pave the way for scaling superconducting quantum devices through modular quantum networks. 
\end{sciabstract}

\title{On-demand quantum state transfer and entanglement between remote microwave cavity memories}

\author{Christopher Axline}
\thanks{Equal contribution}
\email{christopher.axline@yale.edu}
\author{Luke Burkhart}
\thanks{Equal contribution}
\email{luke.burkhart@yale.edu}
\affiliation{Departments of Applied Physics and Physics, Yale University, New Haven, CT 06520, USA}
\author{Wolfgang Pfaff}
\thanks{Equal contribution}
\email{wolfgang.pfaff@microsoft.com}
\affiliation{Departments of Applied Physics and Physics, Yale University, New Haven, CT 06520, USA}
\affiliation{Present address: Microsoft Station Q Delft, 2600 GA Delft, The Netherlands}
\author{Mengzhen Zhang}
\author{Kevin Chou}
\author{Philippe Campagne-Ibarcq}
\author{Philip Reinhold}
\author{Luigi Frunzio}
\author{S.M.~Girvin}
\author{Liang Jiang}
\author{M.H.~Devoret}
\author{R.J.~Schoelkopf}
\email{robert.schoelkopf@yale.edu}
\affiliation{Departments of Applied Physics and Physics, Yale University, New Haven, CT 06520, USA}

\date{\today}
\maketitle
\makeatletter
\def\l@subsubsection#1#2{}
\makeatother

\begin{figure*}[t]
    \input{figures/fig1}
\end{figure*}

\begin{figure*}[t]
    \input{figures/fig2}
\end{figure*}

\begin{figure*}[t]
    \input{figures/fig3}
\end{figure*}

\begin{figure*}[t]
    \input{figures/fig4}
\end{figure*}

The assembly of large-scale quantum machines hinges on the ability to coherently connect individually controlled quantum storage elements.
Quantum networks---wherein small, highly coherent modules can exchange quantum information via propagating photons---present a promising approach to achieve this connectivity \cite{Kimble2008}.
Such networks allow for bottom-up construction of reconfigurable quantum systems, forming a backbone for fault-tolerant modular quantum computers \cite{Jiang2007,Nickerson2013,Monroe2014}. 
A crucial challenge, however, is presented by inefficiencies in the mapping of stored quantum information onto traveling signals as well as those during the subsequent photon transfer.
Primarily because these inefficiencies have so far been large, quantum communication between remote memories has only been achieved probabilistically \cite{Chou2005,Moehring2007,Ritter2012,Hofmann2012,Bernien2013,Roch2014,Narla2016}, requiring local storage of quantum information on long time scales in order for a network to be scalable \cite{Hucul2015}.
Even simple protocols, such as transferring a single quantum bit in a network, have been executed at rates that are orders of magnitude slower than available coherence times \cite{Olmschenk2009,Pfaff2014}.

Direct quantum state transfer, which can be rapid and deterministic, is a desirable scheme for quantum communication within a scalable network \cite{Cirac1997}.
In this protocol, a sending system emits a quantum state as a shaped photonic wavepacket that is then absorbed by a receiving system.
This requires strong, tunable coupling between light and matter, as well as efficient transfer of photons at a shared communication frequency; so far, state transfer in optical networks has been highly probabilistic due to inefficiencies in photon coupling and transfer \cite{Ritter2012}.
In contrast, superconducting microwave circuits can combine low loss with strong coupling.
This platform is well-suited to realize on-demand state transfer, and thus to scale quantum devices in a modular fashion.
To this end, superconducting microwave memories and propagating modes have successfully been interfaced to realize controlled photon emission \cite{Pechal2014,Yin2013,Srinivasan2014,Pfaff2017} and absorption \cite{Wenner2014,Flurin2015,Reed2017} independently.
Due to the difficulty posed by the need for efficient, frequency-matched photon transfer, however, the goal of deterministic quantum communication at a distance has so far remained elusive \cite{Ibarcq2018}.

Here we show an experimental realization of direct, on-demand state transfer and entanglement between separated superconducting quantum memories connected by 60\,cm of coaxial cable.
Following the concept described in the original proposal \cite{Cirac1997}, we release states from a `sender' node and capture them in a nominally identical `receiver' node. 
In our implementation, remote high-$Q$ superconducting microwave cavities constitute the endpoint memories, and the conversion from stationary to propagating mode is performed using an RF-controlled parametric process \cite{Pfaff2017}. 
The receiver uses the same conversion method to capture the propagating signal.
Using this protocol, we demonstrate the transfer of single- and multi-photon cavity states.
We measure the mean fidelity of the transfer of a single-photon-encoded qubit to confirm that the process successfully realizes a quantum communication channel.
We are further able to generate on-demand remote entanglement by half-transfer of a single photon, and do so at a rate that exceeds the energy relaxation rates of the individual memories.
We find that our measured transfer fidelity is limited by photon loss, offering a means for deterministic improvement of transfer fidelity by quantum error correction with multi-photon bosonic codes. 
Taking advantage of the state-independence of the protocol, we demonstrate the feasibility of this correction approach by transferring a qubit in such a multi-photon encoding.

Our experimental implementation is shown schematically in Fig.~\ref{fig:1}A.
Both sender and receiver are 3D circuit quantum electrodynamics (cQED) modules containing long-lived cavity resonators that serve as quantum memories \cite{Axline2016}.
Each module can be understood to contain two orthogonal cavity modes (memory and communication) that are coupled by an artificial atom (Fig.~\ref{fig:1}B).
The communication modes---implemented as on-chip stripline resonators---are strongly coupled to either end of a transmission line.
Realizing on-demand state transfer requires tunable conversion between memory and communication modes within each module, such that (i) the sender emits the state contained in the memory into the transmission line as a wavepacket with a specified temporal profile, and (ii) the receiver absorbs this wavepacket. 
Amplitude- and phase-controlled coupling can be realized through parametric pumping via a single transmon dispersively coupled to both resonators in the module (Fig.~\ref{fig:1}, C and D) \cite{Pfaff2017}.
In particular, we compute the shape of the pumps used in this process so as to best match the temporal profile of the traveling wavepacket \supp.
System parameters enable the effective coupling strengths between memories and the transmission line, $\kappa^\mathrm{s,r}(t)/2\pi$, to be tuned dynamically up to 400\,kHz---much larger than the intrinsic single photon decay rates of the memories, $\kappa_0^\mathrm{s,r}/2\pi <$ 0.4\,kHz \supp. 

Following the original proposal \cite{Cirac1997}, we insert a circulator into the transmission channel, which enforces the directionality of emission from the sender.
The circulator also directs signals reflected off the receiver into an output port, which allows readout of both systems using a single parametric amplifier and heterodyne detection chain. 
While the memory resonance frequencies need not match, efficient transfer requires that the communication modes be close to resonant compared to their bandwidths ($\kappa_\mathrm{out}^\mathrm{s,r}/2\pi \sim $1\,MHz).
To compensate for a small offset in resonance frequency due to variation in sample assembly, we equip the sender with an {\em in situ} mechanical frequency tuning mechanism \supp. 

We begin by characterizing the process by which photons in the sender are emitted, transferred, and absorbed into the receiver memory.
First, we quantify the efficiency of absorption alone by preparing a small coherent state in the sender memory, and then executing the protocol under one of two conditions (Fig.~\ref{fig:2}A).
In one case, we omit the capture pulses and monitor reflection from the receiver. 
Here, the emitted wavepacket is fully reflected and recorded by our heterodyne detector (Fig.~\ref{fig:2}B).
In contrast, if we apply the complete set of pulses, this reflection is strongly suppressed. 
By measuring the relative photon flux at the detector, we determine that the receiver absorbs $(93 \pm 1)\%$ of the energy contained in the incident wavepacket. 

To measure the overall transfer efficiency, we prepare few-photon states and apply both release and capture pulses. 
We measure cavity populations before and after the transfer using photon number-dependent spectroscopy on the transmon, which directly provides the relative populations of the cavity number states \cite{Schuster2007}.
We define the transfer efficiency $\eta$ as the average photon number received divided by the average photon number prepared in the sender. 
Figure \ref{fig:2}C presents the populations of both memories after transferring states with mean photon number $\bar{n} = 1$, from which we calculate an efficiency $\eta = 0.74 \pm 0.03$.
Our experimental scheme is independent of input state, verified by measuring the transfer efficiency of a selection of Fock and coherent states with up to $\bar{n} = 4$ (Fig.~\ref{fig:2}D). 

While this transfer efficiency is high, understanding the origin of process imperfections is critical to select optimal error correction protocols and to correct imperfections in future experiments.
We can identify several factors that contribute significantly to transfer inefficiency: undesired transmon excitation, imperfectly shaped pump pulses, and loss in the transmission path. 

For ideal operation of our protocol, the transmons would remain in their ground states during the transfer. 
However, we observe non-negligible stochastic excitation during the transfer process due to thermalization and pump-induced transitions to higher levels \cite{Bishop2010,Sank2016}. 
Unwanted transmon excitation has two important consequences.
For one, an excitation leads to a shift of the resonator frequencies due to their dispersive couplings to the transmon.
This abruptly changes the transfer frequency-matching conditions, manifesting as off-resonant emission by the sender, or imperfect wavepacket absorption by the receiver.
We estimate these effects to lead to an inefficiency of about $2\%$ for emission, and $6\%$ for absorption \supp.
This effect is thus likely the dominant cause of the measured absorption inefficiency.

Secondly, transmon excitation precludes effective measurement of the cavity state.
In this case, cavity measurement indiscriminately returns `yes' to a query of any photon number.
Excitations thus have the effect of reducing average measurement contrast.
By normalizing our measurement data to correct for this, cavity tomography is implicitly conditioned on the transmon having remained in its ground state.
It is therefore useful to view the transmon excitation probability as a ``failure probability'' of the protocol, i.e, we make the conservative assumption that each excitation masks an unsuccessful transfer. 
The efficiency $\eta$ quoted above is then conditioned on the receiver transmon remaining in the ground state, with success probability $p_\mathrm{s}=0.87 \pm 0.03$. 
The conditioned value can be interpreted as the efficiency that would be measured (i) with a perfectly cold transmon or (ii) by heralding on a transmon measurement after the protocol \supp.
The ``deterministic efficiency'' given the transmon temperature observed here is estimated by the product of the conditioned efficiency and the success probability, $\eta_\mathrm{d} \geq p_\mathrm{s}\times \eta = 0.87 \times 0.74 = 0.64\pm0.03$. 
In the transfer characterization to follow, we present both the directly measured (implicitly conditioned) quantities as well as the estimated deterministic ones.
This deterministic value represents a lower bound on the quantity; since these failure events are assumed to be maximally destructive, this is the worst-case scenario \supp.

Additional contributions to transfer inefficiency come from photon loss in the transmission path, which we estimate adds 15\%, as well as imperfect pulse shapes affecting state release and capture, each with an effect around 2\% \supp. 
We note that the bulk of the described imperfections are not fundamental; in particular, improvements to the transmon equilibrium temperature and thermalization rate as well as parameter engineering to avoid pump-induced higher order transitions \cite{Sank2016} can substantially reduce the inefficiencies resulting from transmon excitation.

The achieved transfer efficiency allows for quantum communication between the sender and the receiver memories.
We confirm this explicitly by transferring an overcomplete set of qubit states in the manifold spanned by the Fock states $\ket{0}$ and $\ket{1}$, and performing Wigner tomography on the receiver (Fig.~\ref{fig:3}A).
Comparing each received state to the ideal state, we determine an average fidelity $\mathcal{F}_\mathrm{avg} = 0.87 \pm 0.04$ (deterministic: $\mathcal{F}_\mathrm{avg,d} \geq p_\mathrm{s}\times\mathcal{F}_\mathrm{avg} = 0.76 \pm 0.04$).
Both the conditioned and deterministic fidelities significantly exceed the classical bound of $\frac{2}{3}$, the maximum attainable fidelity with which one can reconstruct an unknown qubit state using only classical communication \cite{Massar1995}. 

Importantly, the measured fidelity $\mathcal{F}_\mathrm{avg}$ is consistent with that expected ($0.91 \pm 0.03$) from a pure photon loss model using the measured transfer efficiency.
A representation of prepared and received states as vertices of an octahedron on the Bloch sphere (Fig.~\ref{fig:3}B) reveals a systematic shrinkage towards $\ket{0}$ that also appears consistent with photon loss.
More quantitatively, we find that the measured transfer has a process fidelity of 0.95 to this single-source model, bounding the errors not described by photon loss at the 5\% level \supp. 

Our experimental scheme readily enables us to generate on-demand remote entanglement by applying a pump sequence on the sender that releases half of its stored energy (Fig.~\ref{fig:3}C).
If the initial state is a single photon, this results in entanglement between the memory and the emitted radiation \cite{Pfaff2017}.
Subsequent absorption of the wavepacket by the receiver ideally results in the Bell state $\left( \ket{10}+\ket{01} \right ) / \sqrt{2}$ shared between the memories. 
We perform joint tomography following this protocol, revealing non-classical correlations between sender and receiver (Fig.~\ref{fig:3}D). 
Here, the entanglement success probability $p_\mathrm{s,ent}=0.78 \pm 0.04$ is lower than for the state transfer, as success depends on both transmons remaining in the ground state \supp.
The fidelity of the joint state to the ideal Bell state is $\mathcal{F}_\mathrm{Bell} = 0.77 \pm 0.02$ (deterministic: $\mathcal{F}_\mathrm{Bell,d} \geq p_\mathrm{s,ent}\times\mathcal{F}_\mathrm{Bell} = 0.61 \pm 0.02$), confirming the successful generation of entanglement. 
We are able to achieve a net entanglement generation rate of $(140\,\upmu\mathrm{s})^{-1}$ (for fidelity 0.77; equivalently $(110\,\upmu\mathrm{s})^{-1}$ for fidelity 0.61, fully deterministic), limited by the average time it takes to reset the system ($\sim100\,\upmu$s).
This rate exceeds the single-photon loss in either memory ($<(450\,\upmu\mathrm{s})^{-1}$), a strict requirement for scaling up the network size.

Because the infidelity of our state transfer protocol is dominated by errors of a single type---photon loss---the scheme can be improved by selecting an appropriate error-correcting code.
The use of cavity memories grants access to tools already developed for correcting loss in stationary states, such as redundantly encoding a qubit within a larger Hilbert space and using photon number parity as an error syndrome \cite{Ofek2016}.
We choose a simple error-correctable code with minimal overhead, the $L=1$ binomial encoding, which has logical basis states $\left\{\ketL{0}=\ket{2}, \ketL{1}=(\ket{0}+\ket{4})/\sqrt{2}\right\}$ \cite{Michael2016}.
Our transfer scheme is number-state independent, and so with no other modifications we prepare and transmit the cardinal states of this encoding, again measuring the received state with Wigner tomography (Fig.~\ref{fig:4}, A and B).
These states have larger average photon number ($\bar{n} = 2$) relative to the single-photon encoding ($\bar{n} = 0.5$), representing an additional ``overhead''.
From this increased sensitivity to photon loss we predict a mean fidelity of $0.60$ and measure $\mathcal{F}_\mathrm{avg} = 0.54 \pm 0.04$ in this manifold ($\mathcal{F}_\mathrm{avg,d} \geq 0.47 \pm 0.04$) relative to the ideal states.
Though its mean fidelity is lower than that of the single-photon encoding, the binomial encoding permits the use of parity as an error syndrome measurement.

This feature will enable detection and deterministic correction of single-photon loss errors. 
Above some transfer efficiency threshold, an error-corrected qubit would be transmitted with higher average fidelity than a qubit encoded in the single-photon manifold.
Our efficiency ($\eta = 0.74$) exceeds that of this ``break-even'' threshold ($\eta \gtrsim 0.7$), defined as the crossing of simulated mean fidelities in each case (Fig.~\ref{fig:4}C).
Beyond break-even, error correction can overcome the overhead associated with the binomial state encoding.
Error correction is possible using high-fidelity quantum non-demolition (QND) parity measurements \cite{Ofek2016}, which could be effectively realized by adding a dedicated readout channel to each module. 
Along with modest improvements to the release and capture efficiencies, error correction should place the transfer firmly within this advantageous regime.
These error correction concepts can also be extended to improve entanglement fidelity without sacrificing the determinism of the protocol.

In summary, we have realized a high-fidelity, deterministic quantum state transfer protocol between remote microwave cavity memories using tools available in superconducting cQED. 
Importantly, our implementation is capable of transferring both single- and multi-photon quantum states. 
While the use of a multi-photon qubit encoding produces larger overhead and reduced mean fidelity, it brings within reach the striking opportunity to deterministically correct for photon loss.
This achievement taps into the body of work already developed for correcting errors in stationary memories to address the challenge of scalable quantum communication.
The demonstrated ability to generate remote entanglement at a rate exceeding the memory loss rates satisfies an essential requirement for quantum communication and distributed computation.
Entanglement is a critical resource in quantum networks, and its rapid and on-demand generation will enable high-level operations between remote modules such as non-local gates \cite{Gottesman1999} and entanglement distillation \cite{Bennett1996,Deutsch1996}.
Our experimental results thus provide a clear scaling approach towards fault-tolerant modular quantum computing with superconducting circuits.
\section*{Acknowledgements}
The authors would like to acknowledge valuable discussions with C. Zhou, A. Narla, and S. Shankar.
This work was supported by the US Army Research Office (W911NF-14-1-0011). 
C.A. acknowledges support from the NSF Graduate Research Fellowship (DGE-1122492);
L.D.B by the ARO QuaCGR Fellowship;
W.P. by NSF grant PHY1309996 and by a fellowship instituted with a Max Planck Research Award from the Alexander von Humboldt Foundation;
W.P. and P.R. by the US Air Force Office of Scientific Research (FA9550-15-1-0015); 
M.Z. by the US Air Force Office of Scientific Research (FA9550-15-1-0015); 
S.M.G by the NSF (DMR-1609326); 
L.J. by the Alfred P. Sloan Foundation and the Packard Foundation.
Facilities use was supported by the Yale Institute for Nanoscience and Quantum Engineering (YINQE) and the Yale SEAS cleanroom.
R.J.S., and L.F. are founders and equity shareholders of Quantum Circuits, Inc.

\setcounter{equation}{0}
\setcounter{figure}{0}
\setcounter{table}{0}

\renewcommand{\theequation}{S\arabic{equation}}
\renewcommand{\thefigure}{S\arabic{figure}}
\renewcommand{\thetable}{S\arabic{table}}

\clearpage
\newpage
\section*{Supplemental Material\\~}
\section{Experimental system}

\subsection{Experimental hardware}

\begin{figure*}[tp]
    \center
    \includegraphics{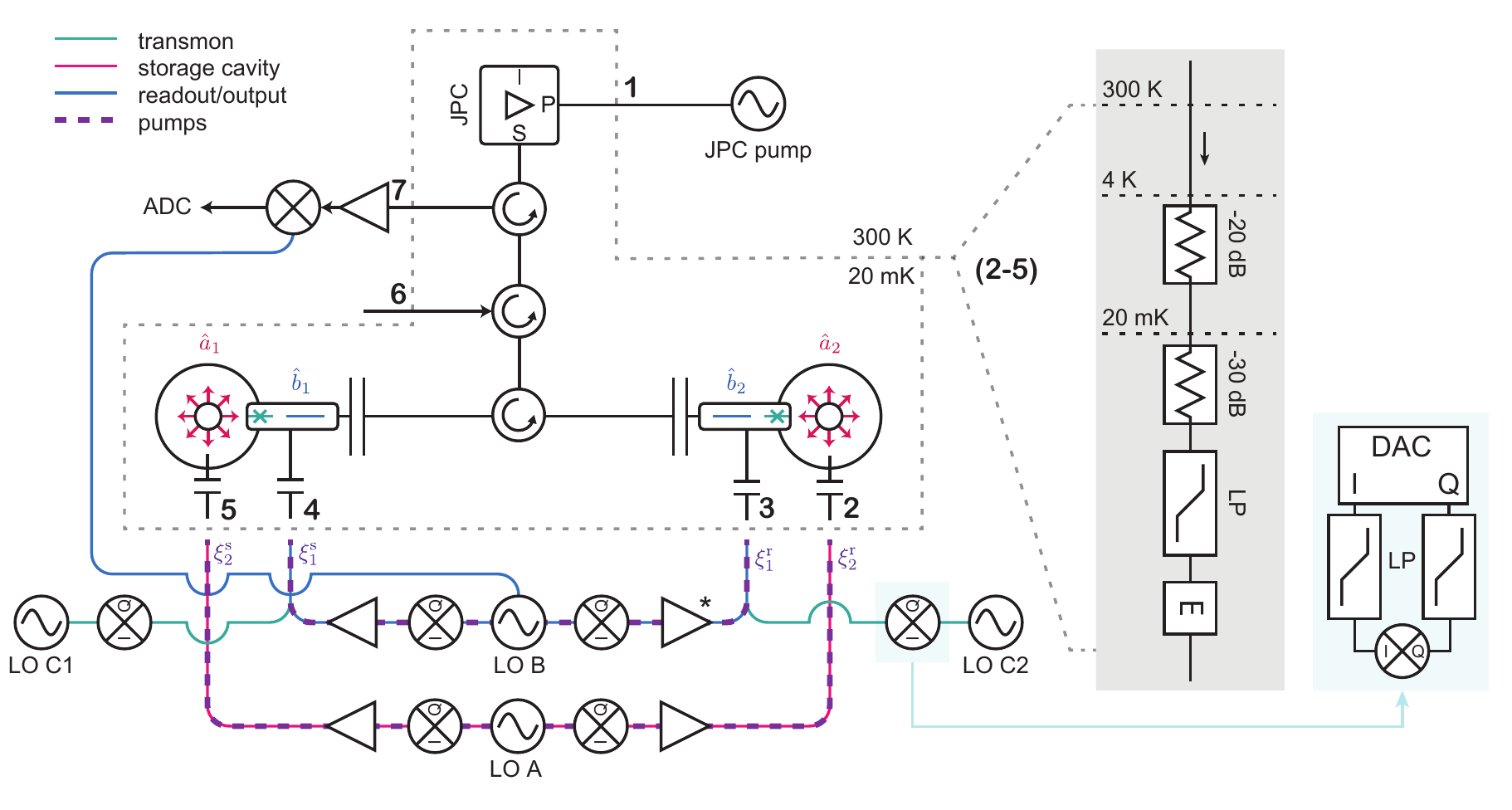}
    \caption{\label{fig:expt_scheme}
    {\bf Schematic of experiment wiring.}
    Microwave signals are IQ-modulated, amplified, and transmitted to the cold stage of the dilution refrigerator ({\em Oxford Instruments Triton 200}) where the sample is cooled to $T \approx 20\,\text{mK}$.
    IQ modulation is performed with an integrated FPGA system with digital-to-analog converter (DAC) outputs ({\em Innovative Integration VPXI-ePC}), and mixed with the LO ({\em Marki Microwave IQ0307LXP and IQ0618LXP} IQ-mixers).
    Inset shows filtering on low-frequency IQ channels.
    The generators acting as local oscillators (LOs) are {\em Agilent N5183A} (LOs A and B) and {\em Vaunix LabBrick LMS-103-13} (LOs C1 and C2).
    Tones with similar frequencies share their IQ channels.
    Cavity and readout input signals are amplified at room temperature ({\em MiniCircuits ZVA-183-S+}). Input amplifiers are followed by fast switches ({\em Hittite HMC-C019}) except for (*) in the diagram.
    Input signals are introduced to the devices through weakly coupled pins (depicted as small capacitors).
    Signals leave the output resonators through strongly coupled pins (depicted as large capacitors) and are amplified by a Josephson parametric converter (JPC) that is pumped continuously by a microwave generator (``JPC pump'', {\em Agilent N5183A}).
    The signal is further amplified at 4\,K ({\em Low Noise Factory LNF-LNC7\_10A}) and at room temperature ({\em Miteq AMF-5F-04001200-15-10P}), mixed down with the output LO ({\em Marki Microwave IQ0618LXP}), and recorded and demodulated by the FPGA system via analog-to-digital converters (ADC).
    Lines entering or leaving the refrigerator carry the following signals: (1) the JPC pump tone, (2--5) sample input tones, (6) a diagnostic tone with which to probe samples in reflection and tune the JPC, and (7) the measurement output line.
    Inset shows attenuation and filtering on lines 2--5, including a low-pass filter below 12\,GHz (LP) and an Eccosorb filter (E).
    Line 6 is similar but with an additional -20 dB of attenuation at the $20\,\text{mK}$ stage, while line 1 omits the low-pass filter.
    }
\end{figure*}

\paragraph*{Setup and signals}
The sample is cooled to $T \approx 20\,\text{mK}$ in a dilution refrigerator.
A wiring diagram that shows how signals are introduced to the systems is depicted in \figref\ref{fig:expt_scheme}.
The transmon modes of each system are addressed by a separate microwave generator acting as a local oscillator (LO), while a shared LO is used for the cavity modes and the output modes; pulses are generated by IQ modulation.
To ensure phase-locking, we generate the pump tones using the same generators used for the (near-)resonant control pulses of the cavity and output mode.

\paragraph*{Quantum-limited amplification}
The output signal is processed by a Josephson parametric converter (JPC) operated in amplification mode \cite{Bergeal2010}.
The amplifier is configured to provide approximately 22\,dB of gain with a bandwidth of approximately 17\,MHz and a noise visibility ratio of 4--5\,dB.
This large signal gain enables high-fidelity single-shot measurements and fast-feedback used for rapid reinitialization of the experiment.

\paragraph*{Sample}
We use two devices of the type described in \cite{Axline2016,Pfaff2017}.
Each includes a high-Q 3D coaxial stub cavity \cite{Reagor2016}, transmon qubit, and output resonator.
Parameters of each mode within the two devices, including frequencies and coherence times, are included in Table \ref{tab:devparams}.
Both samples are within a magnetic shield, while all circulators are outside of the shield.
A detailed description of the Hamiltonian of a single device is given in \cite{Pfaff2017}.

\paragraph*{Output mode-matching}
The cavity and transmon modes of the sender system are far-detuned (relative to their linewidths) from the respective modes in the receiver system. 
The output (or `communication') modes, however, are tuned to closely match:
a low-temperature mechanical micropositioning stage ({\em Attocube ANPz101-A4}) is used to introduce a small superconducting pin through a hole in the side of the device.
The pin perturbs the field of the output mode, lowering that mode's frequency at greater pin insertion.
The pin can be inserted over a range of $5\,\text{mm}$ and the mode frequency can be tuned by $\sim 300\,\text{MHz}$.
Once the resonance condition is met, we disconnect the positioning controller to maintain frequency stability.
We expect that in future versions of modules, fabrication and mounting errors can be minimized such that {\em in situ} tuning will not be necessary.

\begin{figure}[tp]
    \center
    \includegraphics{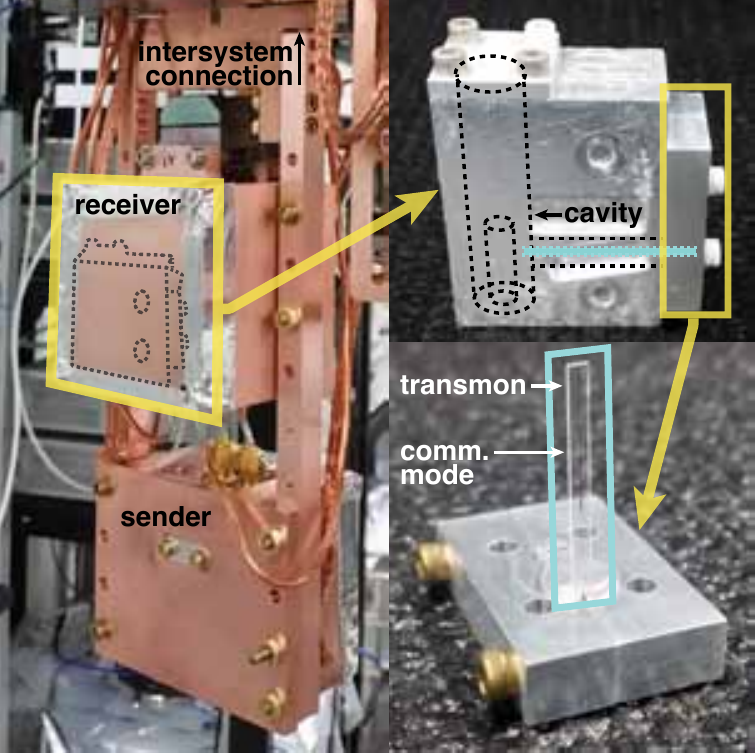}
    \caption{\label{fig:device}
    {\bf Images of experimental devices.}
    {\bf Left:} the two systems are packaged and assembled in close proximity.
    The circulator and lines connecting the two systems are outside of the field of the image.
    {\bf Upper right:} inside the package, dashed lines depict the cavity and tunnel housing the communication mode and transmon.
    {\bf Lower right:} Close-up view of the chip, which is held by a clamp that is affixed to the package.
    }
\end{figure}

\begin{table*}[tb]
    \centering
    \begin{tabular}{l c c c} 
    \hline\hline
    Hamiltonian parameter &  & Sender value (MHz) & Receiver value (MHz) \\
    \hline
    Frequency    & $\omega_{a}/2\pi$      & $4219.3$               & $4269.6$  \\
                      & $\omega_{b}/2\pi$	   & $10031.5$              & $10031.5^{*}$  \\
                      & $\omega_{t}/2\pi$	   & $6156.1$               & $6417.6$   \\
    Cross-Kerr        & $\chiab / 2\pi$     & $-16 \times 10^{-3}$    & $-12 \times 10^{-3}$ \\
                      & $\chiac / 2\pi$     & $-2.86$                & $-2.29$ \\
                      & $\chibc / 2\pi$     & $-2.4$                 & $-2.18$ \\
    Self-Kerr         & $\chiaa / 2\pi$     & $-8 \times 10^{-3}$    & $-5 \times 10^{-3}$ \\
                      & $\chibb / 2\pi$     & $-8 \times 10^{-3}$    & $-6 \times 10^{-3}$ \\
                      & $\chicc / 2\pi$     & $-183.43$              & $-196.17$ \\
    \hline
    Damping parameter & & Sender value ($\upmu$s) & Receiver value ($\upmu$s) \\
    \hline
    Cavity single-photon energy decay time & $T^{a}_{1}$ & $460 \pm 10$       & $770 \pm 10$  \\
    Cavity Ramsey decay time       & $T^{a}_{2R}$        & $102 \pm 3$        & $130 \pm 4$ \\
    Output mode energy decay time  & $T^{b}_{1}$         & $0.14 \pm 0.01$    & $0.11 \pm 0.01$ \\
    Transmon relaxation time & $T^{t}_{1}$  	     & $26 \pm 3$         & $27 \pm 3$ \\
    Transmon Ramsey decay time     & $T^{t}_{2R}$	     & $12 \pm 2$         & $12 \pm 2$ \\
    Transmon Hahn echo decay time  & $T^{t}_{2E}$	     & $15 \pm 2$         & $15 \pm 2$ \\
    \hline
    Steady-state excitation & & Sender value & Receiver value \\
    \hline
    Transmon & $1 - P(g)$        & $0.195$      & $0.209$ \\
    Cavity   & $\bar{n}$     & $0.166$      & $0.172$ \\
    \hline\hline
    \end{tabular}

    \caption{
    \label{tab:devparams}
    {\bf Measured system parameters.}
    Uncertainties of measured Hamiltonian parameters are $< 0.1\%$ except when indicated by fewer significant digits.
    For the cavity and transmon decay times, the uncertainties given are the typical fluctuations observed over the course of one day.
    ${}^{*}$This is the frequency set by tuning and is used during all phases of the experiment.
    }
\end{table*}

\subsection{Basic operation}
\label{sec:operations}

\paragraph*{Transmon measurement}
The dispersive coupling between the transmon and output resonator allows measurement of the transmon qubit state, discriminating between its basis states $\ket{g}$ and $\ket{e}$ with high fidelity \cite{Reed2010}.
Our setup allows for transmission pulses to be applied to both systems simultaneously, allowing discrimination of four joint basis states: $\left\{ \ket{gg}, \ket{ge}, \ket{eg}, \ket{ee} \right\}$.
We can thus measure the sender state, the receiver state, or both states in the same measurement.
We use a $0.7\,\upmu\text{s}$ readout pulse with an average of $\bar{n} \approx 10$ ($5$) photons in the sender (receiver). This allows discrimination of the two basis states in each system with a fidelity $> 0.99$. 
Further error arises from the presence of transmon population in higher states, such as $\ket{f}$.
Readout errors are further limited by transmon state jumps during readout producing a readout fidelity $\geq 0.93$ ($0.93$ for assignment to $\ket{g}$, and $0.98$ for assignment to $\ket{e}$).

\begin{figure}[htp]
    \center
    \includegraphics{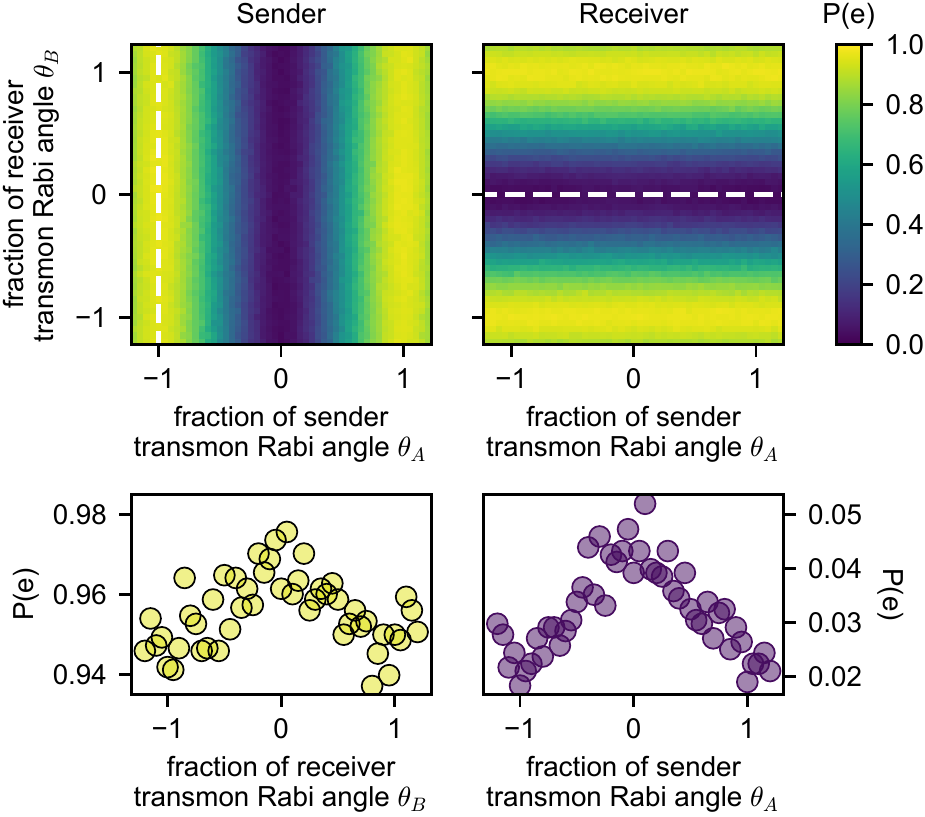}
    \caption{\label{fig:simul_rabi}
    {\bf Transmon measurement independence.}
    Rabi pulses are applied with varying angles (relative to a unit $\pi$-pulse) on the sender and the receiver transmons simultaneously.
    Top: both transmon states are measured as a function of both angles.
    Bottom: Line cuts (corresponding to dashed lines).
    Unwanted correlations are $\leq 2.5\%$.
    }
\end{figure}

\paragraph*{Measurement independence}
We verify the independence of the transmon state assignment by performing a simultaneous Rabi experiment: we apply a variable rotation on both transmons, followed by joint measurement of the whole system.
The results (\figref\ref{fig:simul_rabi}) show that the assigned state of one transmon depends only very weakly on the rotation applied to the other, and that there is minimal crosstalk between transmon control pulses.

\paragraph*{Transmon and cavity cooling}
In equilibrium we find a finite steady-state population in each transmon and cavity, listed in \tabref\ref{tab:devparams}.
The likely reason for this level of population is the presence of non-equilibrium quasiparticles that are poorly mitigated by experiment-stage filtering and control of the magnetic field environment, as well as weak thermalization of the intersystem circulator. 
We initialize each system using a sequence of quantum non-demolition (QND) measurements on the transmon and the cavity \cite{Heeres2017}, as well as by $Q$-switching by parametric pumping \cite{Pfaff2017}.
At the beginning of each experiment run, this yields an average transmon population for the sender (receiver) in $\ket{e}$ of $\sim 0.047$ ($0.052$), and an average cavity population of $\bar{n} = 0.023$ ($0.033$) before the start of each measurement sequence.
We note that this joint cooling protocol finds the full system in its ground state after approximately 10\% of the cooling attempts, and that using modes with reduced equilibrium population could significantly increase the experimental repetition rate.

\paragraph*{State creation and control pulses}
To prepare complex cavity states, we apply pulses at transmon and cavity frequencies that are designed using a numerically optimized control technique \cite{Heeres2017}.
This technique produces the cavity states initialized in the sender within 0.3--1.0\,$\upmu$s and with fidelities typically above 98\%.
It is also used to perform unitary mapping operations.
Coherent states are produced in 40\,ns using calibrated Gaussian displacement pulses on the cavity.

\begin{figure*}[tp]
    \center
    \includegraphics{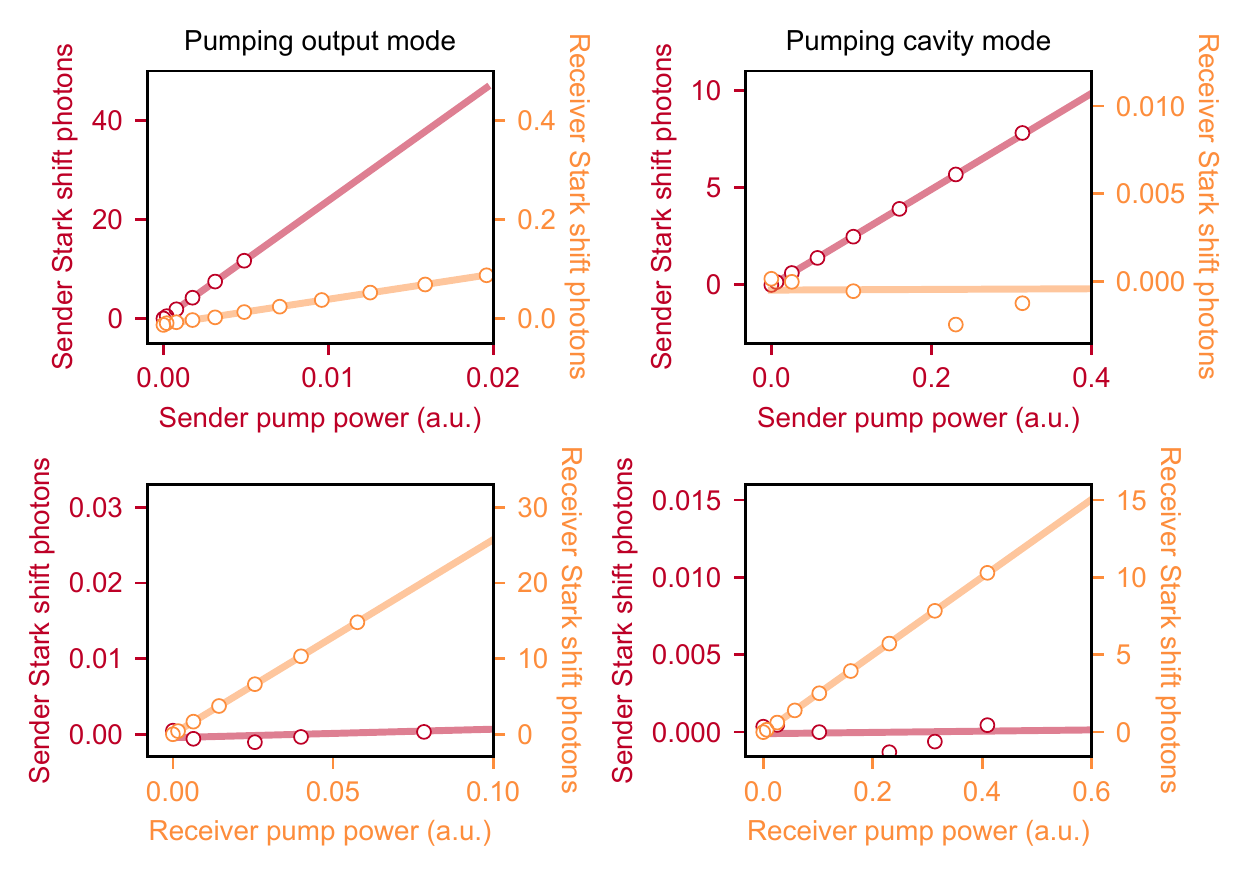}
    \caption{\label{fig:pump_leakage}
    {\bf Pump leakage between systems.}
    We measure the Stark shift (expressed in number of circulating pump photons) on both transmons while applying a single pump to the sender mode (upper row) or the receiver mode (lower row) near either the output/communication mode (left column) or the cavity mode (right column).
    Solid lines are linear fits to the data.
    }
\end{figure*}

\paragraph*{Pump control independence}
To verify that the pump tones applied to one system do not produce unwanted effects on the opposite system, we measure leakage by applying one pump at a constant amplitude and measuring each transmon's Stark shift (\figref\ref{fig:pump_leakage}).
We find that measurable leakage occurs only in the direction permitted by the circulator (from sender to receiver), and only between communication modes (since their frequencies are very nearly matched).
Based on these measurements, we estimate that this leakage contributes to ``miscalibration'' error on each system at the level of $\lesssim 1\%$.

\section{Release and capture protocol} \label{sec:protocol}

\subsection{Calculation of waveforms}

In order to allow the propagating wavepacket to be absorbed by a receiving cavity, both sender and receiver systems must expect the same temporal wavepacket shape.
To separate the problem into two parts, we first specify the shape of this wavepacket, $\bo(t)$, then calculate the pump amplitudes needed at the sender and the receiver to match the coupling rate $g_j(t)$ to this wavepacket, where the system subscript $j = s$ ($r$) denotes the sender (receiver).
When correctly calculated, energy will transfer between the sender, the propagating mode, and the receiver at matching rates.
Here we detail the calculation of the drive amplitudes that make this possible.
Since we treat the two systems individually, the system subscript $j$ will be omitted for clarity wherever possible.

\subsubsection{Shaping the released wavepacket} \label{sssec:shape_release} 
In the rotating frame of the pumps, the Hamiltonian enabling conversion between memory and output modes $a$ and $b$ is
\begin{equation} \label{eq:h_conv}
H_\text{conv}(t) = i(g(t)\hat{a} \hat{b}^\dagger - g^*(t)\hat{a}^\dagger \hat{b})
\end{equation}
with 
\begin{equation} \label{eq:g_of_xi}
g(t) = g(\xi_1(t), \xi_2)
\end{equation}
The correspondence producing the conversion rate $g(t)$ as a function of the two applied pump amplitudes $\xi_1$ and $\xi_2$ is calibrated experimentally (Section \ref{sec:gcal}).
In the lowest-order approximation, $g_j\propto \xi_1^j \xi_2^j$. 

Because the conversion always depends on both pumps, only one of the two pumps needs to vary in time to produce any particular $g(t)$.
For experimental convenience, we designate $\xi_1$ to vary in time.
The pump $\xi_2$ is held constant in time, with a smooth ($\sim200$\,ns) ring-up and ring-down profile.

One significant effect of the application of pumps are ac-Stark shifts that shift the frequency of the modes.
The Stark shifts during the transfer process are a function of both pump amplitudes:
\begin{equation} \label{eq:h_stark}
H_\text{Stark}(t) = \deltaa(t) \hat{a}^\dagger\hat{a} + \deltab(t) \hat{b}^\dagger\hat{b}
\end{equation}
Each Stark shift depends on each pump amplitude:
\begin{equation} \label{eq:delta_of_xi}
\deltaa(t) = \deltaa(\xi_1(t), \xi_2) \qquad \deltab(t) = \deltab(\xi_1(t), \xi_2)
\end{equation}
This shift is calibrated independently (Section \ref{sec:gcal}).
We find very good agreement with the expected dependence, which is linear in the sum of the pump powers:
\begin{equation} \label{eq:delta_starkshifts}
\begin{aligned}
\deltaa(t) &= 2\chiaa |\xi_1(t)|^2 + \chiab |\xi_2|^2 \\ \deltab(t) &= 2\chibb |\xi_2(t)|^2 + \chiab |\xi_1|^2
\end{aligned}
\end{equation}
Having established the dependence of both the conversion rate $g(t)$ and the Stark shifts $\deltaa,\deltab$, we can write down the equations of motion for the modes $\hat{a}$ and $\hat{b}$ of the sender:
\begin{subequations} \label{eq:eom}
\begin{align} \label{seq:eom_a}
&\dot{a}(t) = -g(\xi_1(t), \xi_2)b(t) - i\deltaa(\xi_1(t), \xi_2)a(t)\\ \label{seq:eom_b}
&\dot{b}(t) = g^*(\xi_1(t), \xi_2)a(t) - i\deltab(\xi_1(t), \xi_2)b(t) - \frac{\ko}{2}b(t)\\ \label{seq:eom_c} &b(t) = \bo(t)/\ko
\end{align}
\end{subequations}
Because these equations are linear, we can consider the evolution of the expectation values of the field operators $\hat{a}$ and $\hat{b}$, so we have dropped the operator notation. 
This will also allow us to solve the problem classically, which is computationally simpler compared to a full quantum simulation.

\eqnref\ref{seq:eom_c} is the input-output relation, taking as an assumption that there is no incoming field.
In this case, for a chosen $\bo(t)$ and $\xi_2$, the only undetermined quantities are $a(t)$ and $\xi_1(t)$.
In what follows we suppress the time dependence and the explicit dependence on the static $\xi_2$ for simplicity.

The goal is to eliminate $a$, leaving an equation for $g$ in terms of $b$ that can be solved numerically.
We first note that \eqnref\ref{seq:eom_b} can be written
\begin{equation} \label{eq:elim_a1}
g^*(\xi_1)a = \dot{b} + i\deltab(\xi_1)b + \frac{\ko}{2}b
\end{equation}
with derivative
\begin{equation} \label{eq:elim_a2}
\dot{g}^*(\xi_1)a + g^*(\xi_1)\dot{a} = \ddot{b} + i\dot{\delta}_b(\xi_1)b + i\deltab(\xi_1)\dot{b} + \frac{\ko}{2}\dot{b}
\end{equation}

We can multiply \eqnref\ref{seq:eom_a} by $g^*(\xi_1)$ and substitute it into \eqnref\ref{eq:elim_a2} to write
\begin{multline} \label{eq:elim_a3}
\dot{g}^*(\xi_1)a-|g(\xi_1)|^2b - i\deltaa(\xi_1)g^*(\xi_1)a =\\ \ddot{b} + i\dot{\delta}_b(\xi_1)b + i\deltab(\xi_1)\dot{b} + \frac{\ko}{2}\dot{b}
\end{multline}
Finally, multiplying \eqnref\ref{eq:elim_a3} by $g^*(\xi_1)$ and substituting in \eqnref\ref{eq:elim_a1} yields
\begin{multline}\label{eq:g_of_b}
\left[ \dot{g}^*(\xi_1) - i\deltaa(\xi_1) g^*(\xi_1)\right]\left[\dot{b} + i\deltab(\xi_1)b + \frac{\ko}{2}b\right]\\
-g^*(\xi_1)|g(\xi_1)|^2b   \\
=g^*(\xi_1)\ddot{b} + i\dot{\delta}_b(\xi_1)g^*(\xi_1)b \\
+i\deltab(\xi_1)g^*(\xi_1)\dot{b} + \frac{\ko}{2}g^*(\xi_1)\dot{b}
\end{multline}

\eqnref\ref{eq:g_of_b} is solved numerically to give the correct $\xi_1(t)$ for a given $\bo(t)$.
The initial condition $\xi_1(t=0)$ comes from \eqnref\ref{eq:elim_a1}.
This method has a few important features:
\begin{enumerate}
\item This process inherently accounts for the Stark shifts in two ways.
First, $\xi_1(t)$ will have a phase that varies in time.
This dynamic frequency control ensures that $\bo(t)$ can have a fixed frequency, even when the mode $a$ does not.
Secondly, the amplitude of $\xi_1(t)$ will change in time in a way that accounts for the frequency shift of the output mode $b$: the amplitude will increase to compensate for the fact that the conversion process is effectively off-resonant.
\item By scaling the output field $\bo(t)$ to specify the amount of energy contained therein, we calculate different pump waveforms for full and partial release via the same procedure.
While the equations of motion are linear in $a$ and $b$, \eqnref\ref{eq:g_of_b} is clearly nonlinear in $\xi_1$.
This is why the pulses for full and partial releases are \emph{not} simply scaled versions of one another, even though the released wavepackets $\bo(t)$ \emph{are}.
For this reason, the capture pulse will remain unchanged and independent of the release pulse.
\end{enumerate}

\subsubsection{Wavepacket capture}

The calculation of the pump waveform required to capture the propagating wavepacket is very similar to the above.
The equations of motion for the receiver are
\begin{subequations} \label{eq:eom_catch}
\begin{align} \label{seq:eom_a_catch}
\dot{a}(t) &= -g(\xi_1(t), \xi_2)b(t) - i\deltaa(\xi_1(t), \xi_2)a(t)\\ 
\begin{split} \label{seq:eom_b_catch}
\dot{b}(t) &= g^*(\xi_1(t), \xi_2)a(t) - i\deltab(\xi_1(t), \xi_2)b(t) \\
&\quad{}- \frac{\ko^r}{2}b(t) + \sqrt{\ko^r}\bi^r(t)
\end{split}\\
b(t) &= \bo^r(t)/\ko^r + \bi^r(t)/\ko^r \label{seq:eom_c_catch}
\end{align}
\end{subequations}
which is identical to \eqnref\ref{eq:eom}, with the difference than there now exists an input field term $\bi^r(t)$.
For clarity we now restore the superscripts $s$ and $r$ for sender and receiver, respectively.
To calculate the capture waveform, we specify that this input field has the shape of the released wavepacket: $\bi^r(t) = \bo^s(t)$, and that the field reflected off the receiver is zero: $\bo^r(t)=0$, which corresponds to perfect absorption.
Taken together, these constrains imply
\begin{subequations} \label{eq:eom_catch_reduced}
\begin{align} \label{seq:eom_a_catch_reduced}
&\dot{a}(t) = -g(\xi_1(t), \xi_2)b(t) - i\deltaa(\xi_1(t), \xi_2)a(t)\\ \label{seq:eom_b_catch_reduced}
&\dot{b}(t) = g^*(\xi_1(t), \xi_2)a(t) - i\deltab(\xi_1(t), \xi_2)b(t) + \frac{\ko^r}{2}b(t) \\ \label{seq:eom_c_catch_reduced} 
&b(t) = \bo^s(t)/\ko^r
\end{align}
\end{subequations}
which looks just like \eqnref\ref{eq:eom}, but with the sign of $\ko$ changed.
The procedure for obtaining $\xi_1(t)$ is the same as for the sender.
The only major difference is that this equation is solved in reverse, with the final condition specifying the occupation of $a$ at the end of the protocol.
This corresponds to the fraction of incoming energy that is absorbed ($\eta^\text{(r)}_\text{trunc}$, details to follow).
Increasing this fraction corresponds to increasing the pump strength beyond what is achievable in our system.

Importantly, the capture waveform is the same for both full and partial release; this is due to the linearity of the equations of motion (\eqnref\ref{eq:eom_catch}) on $\bi^r(t)$; in other words, the capture is state-independent.
Therefore, the capture waveform depends only on the shape of the incoming wavepacket, not its amplitude.

\subsubsection{Choice of wavepacket shape}
The envelope of the wavepacket $|\bo(t)|$ is an arbitrary choice, up to constraints on the bandwidth of the conversion process.
For experimental convenience, we choose $|\bo(t)| \propto  1 - \cos^2\left(\frac{\pi t}{T}\right) $, where $T=6\,\upmu \mathrm{s}$ is the total transfer time.
We find empirically that this smooth shape reduces the maximum pump amplitudes required for a given transfer time as compared to other shapes tested.
The frequency of the wavepacket is also free to be varied.
We choose a frequency $\sim 1\, \mathrm{MHz}$ below the static frequencies of the communication modes, to account for Stark shifts (which are always negative) while the pumps are applied.


\subsection{Calibration}
\label{sec:gcal}

For the computation of the correct waveforms we need to supply empirical values of $g(t)$ and the Stark shifts, both as a function of the pump strengths and frequencies that are applied.
The value of $g$ is estimated from the rate with which photons leave the storage cavity when pumps are applied.
The Stark shifts can be measured directly from spectroscopy while applying the pumps, as in \figref\ref{fig:pump_leakage}.
We generally follow the protocols introduced in \cite{Pfaff2017} to perform these calibration.

\subsection{Time-dependence measurement}

\paragraph*{Simulation}
Through the numerical calculation of pulse shapes $\xi_1$ and $\xi_2$ on the sender and receiver (Section \ref{sec:protocol}), we know the ideal intracavity field expectation values $\hat{a}, \hat{b}$ in each system as a function of time.
These are directly related to the average photon number we would expect to measure at that time, $\bar{n} = a^2$.

\begin{figure}[tp]
    \center
    \includegraphics{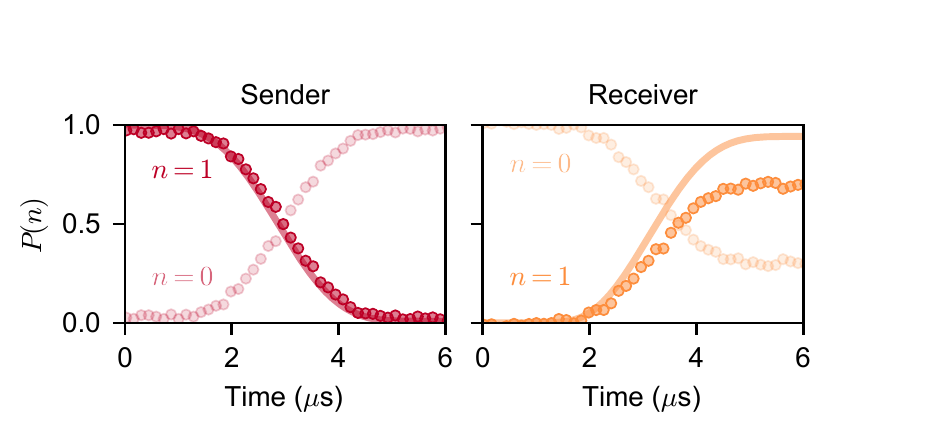}
    \caption{\label{fig:timedep}
    {\bf Cavity populations over time.}
    We prepare and transfer the Fock state with $n=1$.
    Populations for the lowest two cavity number states are measured in both cavities as a function of time.
    The measured shape of the trajectory in the receiver reflects the inefficiency of the transfer.
    }
\end{figure}

\paragraph*{Measurement}
We measure the photon number present in the sender and the receiver after truncating the protocol prematurely: the non-dynamic pulses, $\xi^{(s)}_2$ and $\xi^{(r)}_2$, are turned off over 200\,ns at a swept measurement time $T$.
We then measure the population $a^{(s)}(T)$ and compare with the calculated value.
The results from these measurements are in good agreement with the expected values (\figref\ref{fig:timedep}).

\section{Methods, calibrations, and analysis} \label{sec:methods}

\subsection{Storage cavity measurement}

\begin{figure}[tp]
    \center
    \includegraphics{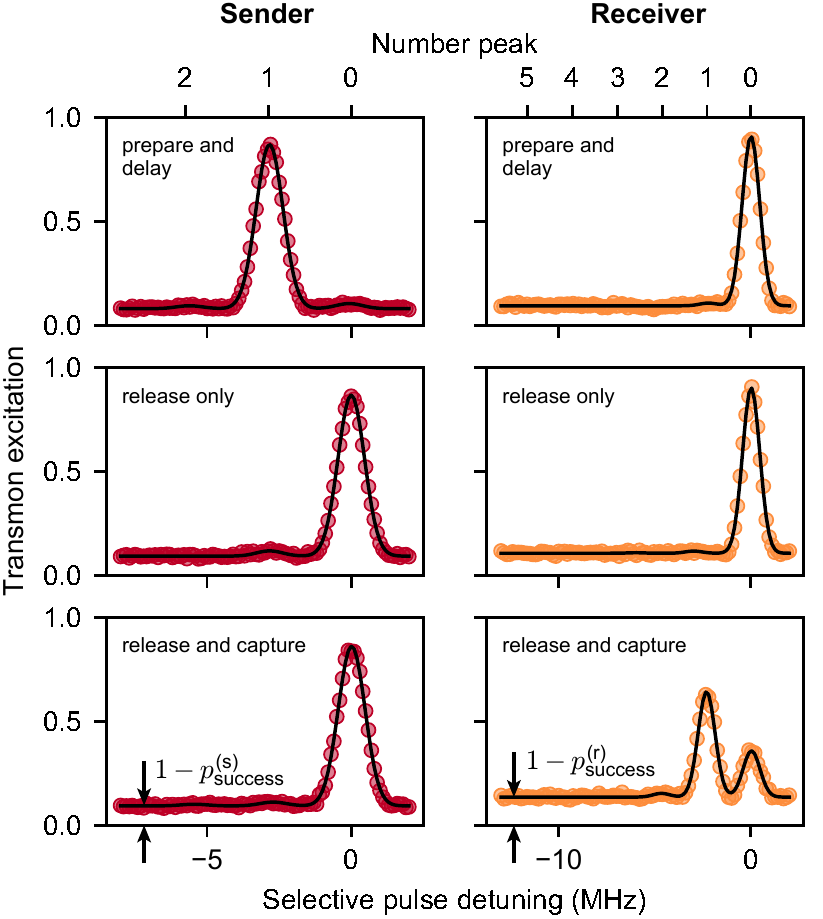}
    \caption{\label{fig:npeakspec}
    {\bf Spectroscopy to determine transfer efficiency.}
    We execute the transfer protocol after preparation of a Fock state with $\ket{1}$.
    The transmon spectrum yields the population of cavity number states.
    Measurement follows different combinations of pumps on the two systems (prepare and delay: the state is prepared in the sender and 6\,$\upmu$s elapse before measurement; release only: pulses are played on the sender but not the receiver; full release and capture: the standard transfer protocol is run).
    Additionally, these data illustrate the rise in background excitation level associated with pumping either system, related to the success probabilities $p^\text{(s,r)}_\text{success}$.
    }
\end{figure}

\paragraph*{Cavity spectroscopy}
Dispersive coupling between transmon and storage cavity allows readout of the cavity state by mapping a cavity property, such as population in a given number state, onto the transmon state \cite{Schuster2007}.
Cavity measurement fidelity is then given by the combined fidelities of the mapping and the transmon readout.
The relative population of each cavity number peak is obtained directly from the transmon spectrum, which depends on the photon number in the cavity.
The spectrum is fit to a series of Gaussians (with total area normalized to one), which directly yields the relative photon number occupations when the transmon is in its ground state.
The raw spectroscopy data producing the relative populations in \figref2C of the main text are shown in \figref\ref{fig:npeakspec}.

\paragraph*{Wigner tomography}
Wigner tomography on a single cavity is done through a series of displaced parity measurements \cite{Haroche2006,Sun2014}.
In our systems, the parity is mapped onto the transmon by applying a set of selective $\pi$-pulses simultaneously, on either the even or odd number-peaks, up to $\bar{n}=25$.
Taking the difference of the even and odd mapping sequences, we obtain a value proportional to parity (and therefore to the Wigner function), independent of the initial transmon state.
To compensate for finite $\pi$-pulse and measurement contrast, we then normalize the resulting data so that it integrates to unity over all phase space, leading to a physical Wigner function.
Displacements are typically performed up to $\alpha = 2.5$, resulting in Wigner functions that should capture $>99\,\%$ of the energy contained in the mode for our states with $\bar{n} \leq 2$.

\subsection{Transfer efficiency analysis} \label{sec:loss}
The combined efficiency of the transfer process can be obtained directly by measurements before and after the transfer protocol, as described in the main text. We provide a breakdown of the different sources of inefficiency in \tabref\ref{tab:loss_params}.

\begin{table*}[tb]
    \centering
    \setlength{\tabcolsep}{8pt}
    \begin{tabular}{p{0.3\linewidth} c c c p{0.3\linewidth}}
    \hline\hline
    Category &  & Value & Uncertainty & Notes \\
    \hline\hline
    \emph{Measured values} \\
    Unreleased energy fraction & $n_\text{remain}^\text{(s)}$  & $0.033$   & $0.005$ & \\
    Uncaught energy fraction & $n_\text{reflect}^\text{(r)}$  & $0.068$   & $0.005$ & \\
    Transmission loss & $\eta_\text{tx,meas}$ & $0.80$ & $0.15$ & \\
    \hline
    {\bf Total efficiency} & $\eta$ & $0.74$ & $0.03$ & \\
    \hline\hline
    \emph{Modeled values} \\
    Truncation of sender pulse              & $\eta_\text{trunc}^\text{(s)}$        & $0.99$        & -                 & specified in pulse creation \\  
    Sender transmon excitation              & $\eta_\text{excite}^\text{(s)}$       & $0.98$        & $0.01$            & \\   
    Calibration of sender pulse             & $\eta_\text{miscal}^\text{(s)}$       & $0.98$       & $0.005$            & \\
    {\bf Release efficiency}                & ${\bf \eta_\text{release}}$           & ${\bf 0.95}$  & $0.01$            & \\
    \hline
    {\bf Transmission line loss}            & ${\bf \eta_\text{tx}}$                & ${\bf 0.85}$  & $0.04$            & estimation from ``missing energy'' \\
    \hline
    Truncation of receiver pulse            & $\eta_\text{trunc}^\text{(r)}$        & $0.95$        & -                 & specified in pulse creation \\
    Truncation of receiver pulse, revised   & $\eta_\text{trunc,corr}^\text{(r)}$   & $0.99$        & $0.005$           & \\
    Receiver transmon excitation            & $\eta_\text{excite}^\text{(r)}$       & $0.94$        & $0.02$            & \\   
    Calibration of receiver pulse           & $\eta_\text{miscal}^\text{(r)}$       & $\geq 0.99$   &                   & \\
    {\bf Capture efficiency}                & ${\bf \eta_\text{capture}}$           & ${\bf 0.92}$  & $0.02$            & \\
    \hline\hline
    \end{tabular}

    \caption{
    \label{tab:loss_params}
    {\bf Summary of transfer losses.}
    A break-down of loss mechanisms thought to contribute to the transfer efficiency.
    For some categories the individual constituent efficiencies are measurable, while for others only the total loss can be measured, and the components must be deduced.
    The model is designed such that the product of the three category subtotals, $\eta_\text{release} \times \eta_\text{tx} \times \eta_\text{capture}$, equals the measured total efficiency $\eta$.
    }
\end{table*}

\subsubsection*{Release and capture efficiency}

\paragraph*{Truncation}
Amplitude and time constraints of the transfer process drives are imposed by the physical implementation of the experimental hardware.
With finite time and amplitude constraints on the conversion rate $g(t)$, releasing and capturing the entirety of the energy in the sender is not possible.
To limit the applied pump power (and thus the amount of heating induced to the system), we implemented the full transfer process specifying waveforms that are expected to release the fraction $\eta^\text{(s)}_\text{trunc} = 0.99$ of the sender state, and capture $\eta^\text{(r)}_\text{trunc} = 0.95$ of the energy in the receiver.



\begin{figure}[tp]
    \center
    \includegraphics{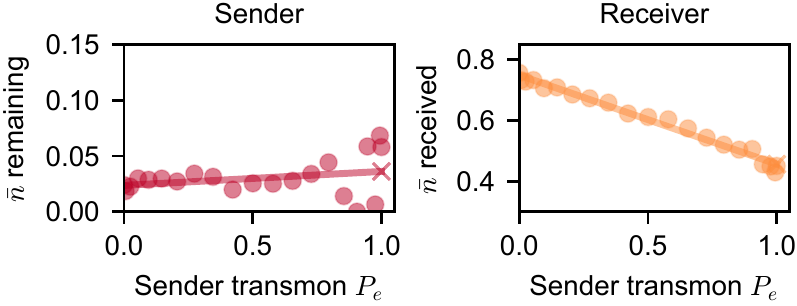}
    \caption{\label{fig:transmon_excite_spec}
    {\bf Transmission inefficiency from sender transmon excitation.}
    We measure both cavities as a function of sender transmon excitation probability.
    ``$\bar{n}$ remaining'' in the sender cavity is inherently conditioned on the sender transmon being in its ground state. The energy remaining appears therefore unaffected.
    The received number of photons is inversely proportional to the sender transmon excitation, reflecting that the state has not been released with proper matching.
    }
\end{figure}

\begin{figure}[tp]
    \center
    \includegraphics{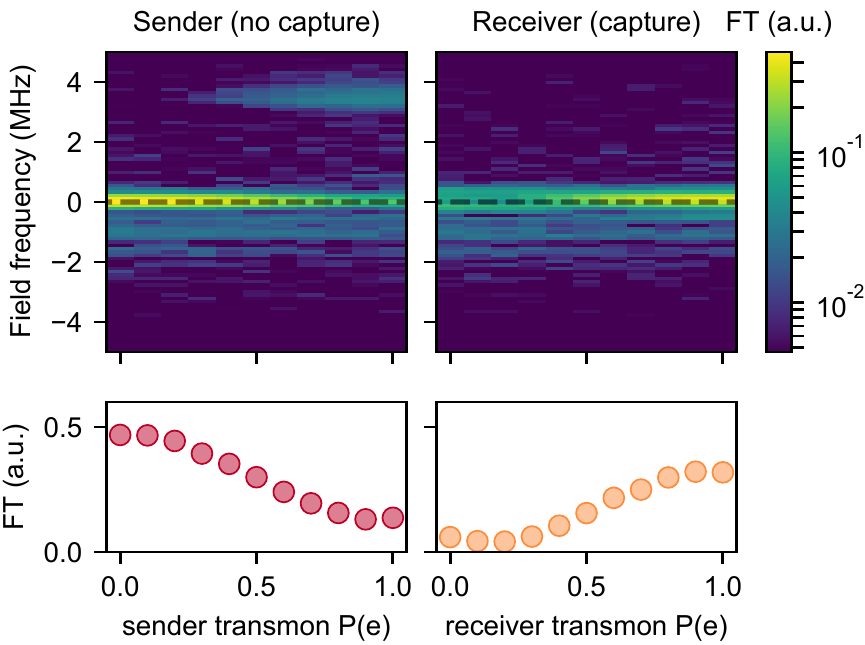}
    \caption{\label{fig:transmon_excite_field}
    {\bf Effect of transmon excitation on the wavepacket.}
    We vary the excitation of the sender transmon (left) or receiver transmon (right), and release a coherent state with $\alpha = 1$ from the sender. During release, we monitor the field leaving the system.
    The magnitude of the Fourier transform (FT) is shown.
    {\bf Left:} The capture pulse is omitted to monitor the released wavepacket.
    A cut taken at the nominal propagating frequency (dashed line) indicates a decrease in emitted field at the transmission frequency.
    At large excitations, field appears to be emitted around 3.5\,MHz above the nominal frequency.
    {\bf Right:} The capture pulses are played, and the reflected field is observed.
    At the nominal propagating frequency, a line cut (dashed) shows that reflected field intensity increases with increased receiver excitation probability (lower).
    The reflected energy received at maximal receiver excitation is 69\% of that received from a normal release with no capture attempt.
    }
\end{figure}

\paragraph*{Sender transmon excitation}
During the transfer, we observe the sender transmon excitation from 3.1\% to 9.6\%, a process from which we approximate a worst-case transfer inefficiency of $\sim 6\%$.
We perform a control experiment, populating the sender transmon before executing the transfer, which allows us to refine our approximation.
A fully-excited sender transmon still transfers somewhat successfully (\figref\ref{fig:transmon_excite_spec}), introducing an inefficiency of only $\sim 39\%$.
We thus revise our worst-case estimate upward, calculating a total efficiency from sender transmon excitation $\eta^\text{(s)}_\text{excite} = 1 - (0.06 \times 0.39) \approx 0.98$.
Our understanding of the effect of transmon excitation on shape/frequency mismatch is improved by monitoring the emitted wavepacket as a function of excitation (\figref\ref{fig:transmon_excite_field}).

\paragraph*{Receiver transmon excitation} 
From the data in \figref\ref{fig:transmon_excite_field} we infer that full excitation of the receiver transmon leads to reflection of $\sim 69\%$ of the incident energy.
Combining this with the excitation of the transmon---increasing from 4.5\% to 13.5\% during the transfer---yields an approximate maximum capture efficiency of $\eta^\text{(r)}_\text{excite} \approx 0.94$.
Taking into account pulse truncation (see above) gives an expected capture efficiency of 0.89.
This is in reasonable agreement with the observed reflection of $6.8\%$ of the incident wavepacket shown in \figref2B of the main text.
The remaining mismatch could be explained by a better value of truncation efficiency than expected due to experimental fine-tuning, $\eta^\text{(r)}_\text{trunc,corr} = 0.994$; this quantity cannot be directly measured in our experiment.

\paragraph*{Miscalibrations} 
Any unidentified source of inefficiency could be due to imperfectly calibrated pulses or uncertainty in the system parameters.
Assuming that missing factors in the capture or release error budgets arise from miscalibration error, we estimate an upper bound.
For the sender, this gives an inefficiency of $\eta^\text{(s)}_\text{miscal} \approx 0.98$.
For the receiver, miscalibrations seem to yield a slightly better capture efficiency than expected (see above); this leaves no gap in the error budget for miscalibration.
Other possible sources of uncertainty include higher-order nonlinearities and frequency-dependent dissipation, which we have not considered in our model.

\subsubsection*{Transmission loss}
\begin{figure}[htp]
    \center
    \includegraphics{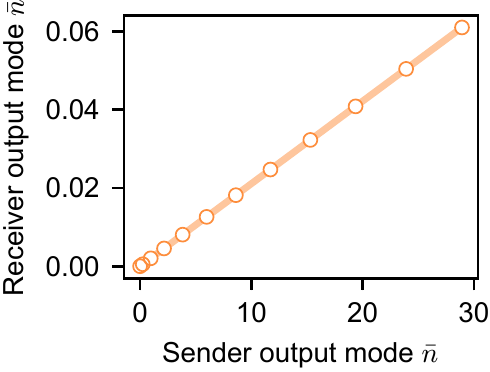}
    \caption{\label{fig:tx_loss}
    {\bf Estimation of transmission loss.}
    We prepare a steady-state population of photons in the sender and measure the resulting population in the receiver by means of the Stark shift.
    This measurement allows us to estimate the fraction of photons transmitted from sender to receiver.
    The line is a linear fit.
    }
\end{figure}

The energy dissipated in the transmission line is difficult to measure directly in our experiment.
We can, however, assume that any `missing' energy comes from loss; this gives a value for the transmission line efficiency of $\eta_\text{tx} = 0.85 \pm 0.04$.
We perform the following control experiment to corroborate this value: we apply a constant tone detuned from the output resonator of the sender, and measure the Stark shift of both sending and receiving communication modes ($\hat{b}^\text{s,r}$) at variable amplitudes of the tone (\figref\ref{fig:tx_loss}).
From these Stark shifts and knowledge of the readout-transmon cross-Kerr $\chibc$ and output rate $\ko$, one can calculate the loss between systems in a way that is independent of the sending and receiving portions of the transfer protocol.
One expects to receive
\begin{equation}
    \label{eq:tx_loss}
    \bar{n}_{r} = \bar{n}_{s} \kappa_s \kappa_r \eta_\text{tx,meas} / ((\kappa_r/2)^2 + \delta_r^2)
\end{equation}
for a transmission loss $\eta_\text{tx,meas}$ between the two communication resonators, and a detuning $\delta_r$ of mode $\hat{b}^{r}$ relative to the drive.
We perform these measurements at several detunings and extract a value of $\eta_\text{tx,meas} = 0.80 \pm 0.15$.
This agrees well with the ``missing energy'' estimation of $\eta_\text{tx}$ above.

\subsection{Cavity state analysis} \label{sec:interpret}

\begin{figure}[htp]
    \center
    \includegraphics{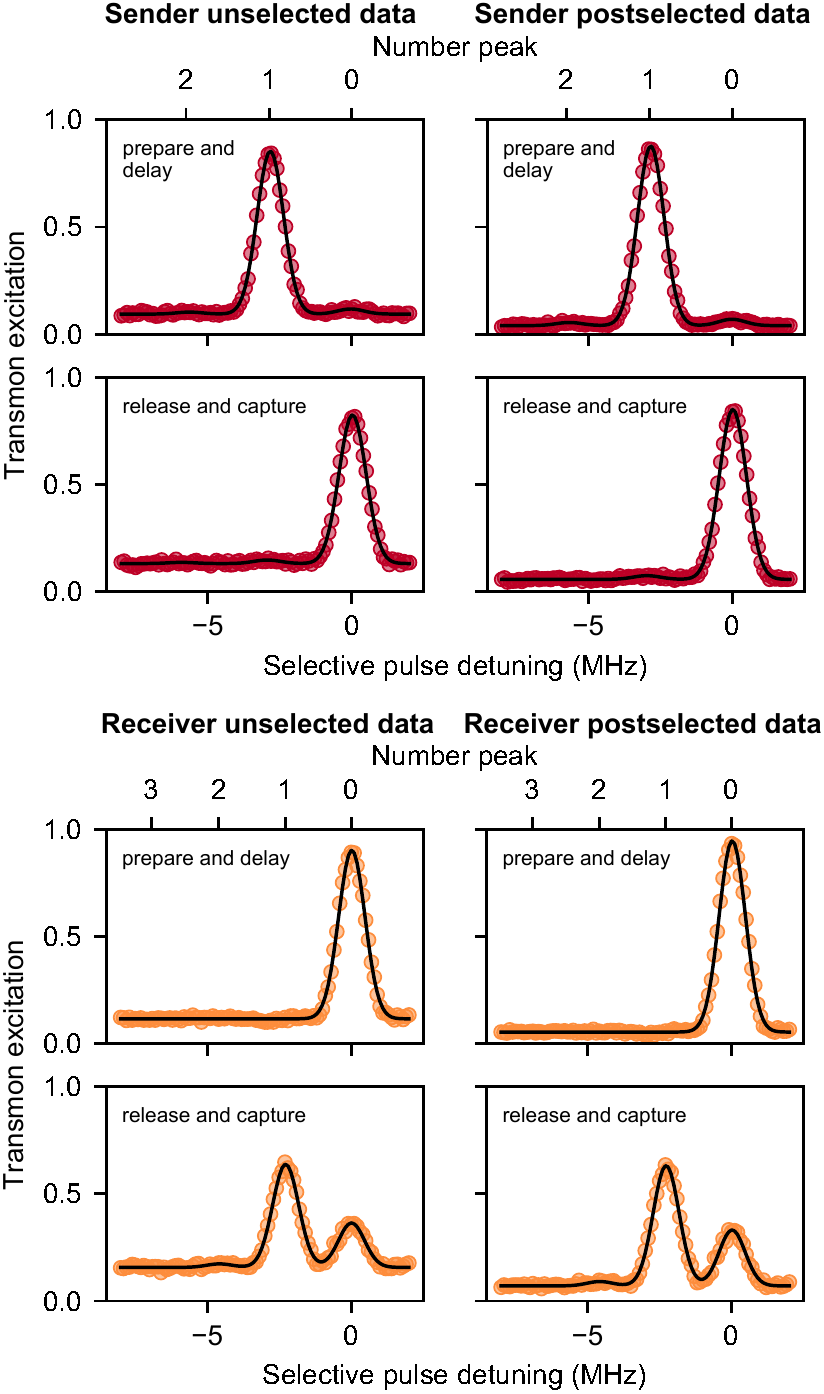}
    \caption{\label{fig:herald}
    {\bf Comparison with post-selected efficiency.}
    Transmon spectroscopy is performed with an added conditioning measurement of transmon state following the transfer process.
    All of the data (left column) or only data post-selected on the conditioning measurement (right column) are shown after preparation (top row) or after release and capture (bottom row).
    Black lines indicate fits from which relative photon number populations are obtained.
    While background offsets are clearly lower in the post-selected data, the extracted relative populations are identical within the range of uncertainty.
    In particular, the measured transfer efficiency, or ratio between average prepared and received photon numbers, changes by $< 1\,\%$.
    }
\end{figure}

\paragraph*{Heralding}
As discussed in the main text, measurements using the receiver transmon have the effect of selecting on that transmon being in the ground state. 
In this case, data that is effectively ignored by normalization conveys no information.
We confirm this by performing a spectroscopy experiment, measuring the transmon state after the transfer process.
We then post-select on the measurement results in each system when the transmon is found in the ground state.
Within fit and measurement error, we find identical photon number content in the unselected (including all data) and post-selected cases (\figref\ref{fig:herald}).
In other words, the average efficiency $\eta$ does not change whether a measurement to detect transmon state is performed, and the discarded measurements contain no relevant information.
Therefore, a measurement of the transmon ground state can be thought of as heralding on a successful state transfer (a transfer not precluded by an endpoint failure, yet still subject to miscalibrations and intermediate photon loss).
All data presented in this document are naturally ``conditioned'' as such, unless otherwise stated.

\paragraph*{Deterministic efficiency and fidelity bounds}
As stated in the main text, the deterministic estimates for the efficiency ($\eta_\mathrm{d}$) and fidelities ($\mathcal{F}_\mathrm{avg,d}$ and $\mathcal{F}_\mathrm{Bell,d}$) are calculated as the product of the conditioned quantity and the success probability.
In each case, this quantity is a lower bound.
For the efficiency, we assume that no energy was captured in the event that the transmon is found excited.
In reality, some energy was likely absorbed in the time before the transmon becomes excited.
For each fidelity bound, we take the worst-case assumption: that the entire system leaves the codespace entirely in an unrecoverable way, and these events contribute zero fidelity to the average.
In fact, if the failure completely destroys the information but leaves the system in a state which can be detected and reset (for example, the transmon in $|e\rangle$), then these events can result in an average fidelity $0.5$ by resetting the system to an arbitrary state within the codespace.
The same is true for the entanglement (but with fidelity $0.25$, which is the best that can be prepared with local resources).
Because some of the failure events are in principle recoverable in this fashion, we quote our worst-case fidelity as a lower bound.

\paragraph*{MLE reconstruction}
Maximum-likelihood estimation (MLE) reconstruction is performed on measured Wigner tomography data to return the density matrix found to have most probably produced the data.
We use a numerical, iterative, convex optimization algorithm.
The algorithm is supplied with normalized (physical) Wigner tomography data, and the optimal reconstructed density matrix is constrained to remain physical throughout the minimization procedure.
Through systematic simulation as well as comparison with other reconstruction methods, we find that this reconstruction technique misrepresents the collected data $<1\,\%$.
The same optimization algorithm is applied to joint cavity measurement operator data in order to reconstruct the density matrix for an entangled state.

\paragraph*{Error estimation}
Uncertainties in values taken from Wigner function reconstructions, such as state fidelities, arise from one of several errors: miscalibrated displacements, improper normalization, or effects of unwanted thermalization.
The amplitude associated with displacing a state by, for example, $\alpha=1$, can be calibrated in several ways.
These calibration values and their relative uncertainties produce a total uncertainty around 1\%.
Variable transmon excitation leads to a normalization factor that can be state-dependent (depending on the preparation method), and so normalization must be performed on every state individually.
Uncertainty in the normalization process arises from the action of integrating over noisy data, and leads to error of several percent.
Thermal effects can alter the shape and spread of the Gaussian form of the vacuum Wigner function.
Each of these errors, in fact, are most pronounced when measuring the vacuum state, since this can easily lead to a nonphysical density state reconstruction.
We analyze the effect of these errors on well-defined vacuum states in each cavity to produce a total, conservative uncertainty associated with Wigner function measurement and reconstruction equal to 4\%.

\subsection{Entanglement generation by partial transmission}
\paragraph*{Interaction between cavity and traveling mode}
The interaction between the cavity and the transmission line can be described by an effective beam splitter interaction (\eqnref\ref{eq:bs_theta}), where the conversion strength and time determine the coupling $\theta$ between the modes \cite{Pfaff2017},
\begin{equation} \label{eq:bs_theta}
H_\text{bs} = \frac{-i \theta}{2} (\hat{a}_1^\dag \hat{a}_2 - \hat{a}_1 \hat{a}_2^\dag)
\end{equation}
In particular, releasing half the energy stored in the cavity initially corresponds to a `50:50 beam splitter' with $\theta = \pi/2$.
The pulse required to realize this can be calculated precisely by specifying that one-half of the prepared state remains (in \eqnref\ref{eq:eom}).

\paragraph*{Correlation measurements and state reconstruction}
In each encoding we perform the following experimental sequence.
We first release half the energy of the prepared state, capturing it in the receiver.
Immediately after switching off the pumps, we apply a mapping pulse that rotates the state of one or both cavities into a particular basis.
Within that basis, we then measure the likelihood of finding the cavities in a particular joint photon-number state by applying selective-$\pi$ pulses on each transmon and reading out.
In particular, for each module we measure after $\pi$ pulses selective on the $n=0$ and $n=1$ peaks, as well as with no $\pi$ pulse (to obtain the background transmon excitation probability).
These $3\times3=9$ probabilities are combined to produce four values for each rotation: the probability of finding joint photon numbers $00$, $01$, $10$, or $11$ (the diagonal elements of the density matrix in this joint basis).
We choose three basis rotations for each cavity: $\{I,Y(\frac{\pi}{2}),X(\frac{\pi}{2})\}$, leading to measurements in the $\{z,x,y\}$ bases, respectively.
This produces correlation probabilities in $3\times3=9$ joint bases, enough to reconstruct the full joint state.
The data for each rotation are supplied to a maximum likelihood estimation (MLE) reconstruction program to produce the density matrix describing the joint system with the largest likelihood.
The Pauli bars in \figref3D are then calculated from this reconstructed density matrix.

\begin{figure}[htp]
    \center
    \includegraphics{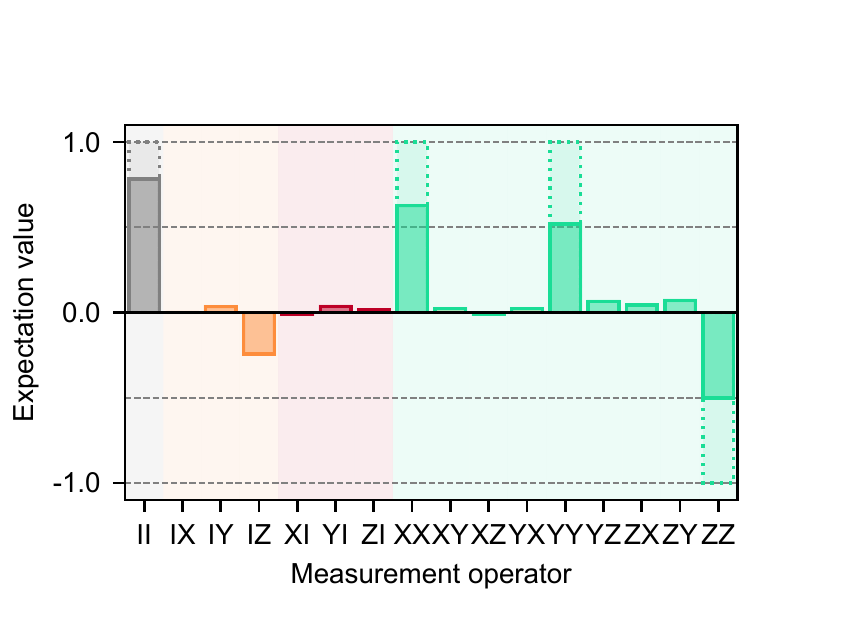}
    \caption{\label{fig:pauli_unconditioned}
    {\bf Entanglement unconditioned by successes.}
    The values measured in \figref2D are calculated with consideration of the joint success probability of the entanglement generation procedure.
    This produces inferred expectation values of two-qubit measurement operators that correspond with the `unconditioned' case, the values we would expect to measure if the transmons did not leave the observable measurement space.
    The measured value associated with the `II' operator, 0.78, thus represents the reduction from unity of this joint success probability compared to the `conditioned' case presented in the main text.
    The fidelity of this unconditioned density matrix to the maximally entangled state is 61\%.
    The scaled measured values (shaded) are overlaid on the ideal values (dashed).
    }
\end{figure}

\paragraph*{Success rate conditioning} \label{sec:ent_succ_cond}
The measurement of joint cavity operators is affected by the success probability of each transmon in a way similar to that of spectroscopy and Wigner tomography measurements (Section~\ref{sec:interpret}).
The data presented in \figref3D of the main text are thus naturally conditioned on the case where both transmons remain in the ground state following the half-transfer and before the measurement is performed.
The unconditioned values (and resulting entanglement metrics) can be determined by accounting for the probability with which each transmon becomes excited or the system leaves the observed Hilbert space (\figref\ref{fig:pauli_unconditioned}).
As opposed to other measurements throughout this work, the excitation of \emph{both} systems factors in.
To estimate the unconditioned entanglement measurement operator expectation values, we add a extra dimension to the reconstructed density matrix representing the measured joint cavity state.
This dimension represents a space outside of the observable space.
It is populated according to the product of the two transmon excitation probabilities.
The density matrix is then renormalized and truncated to exclude the `unobservable' dimension, producing a density matrix reasonably estimating what we believe an unconditioned measurement would produce.

\paragraph*{Entanglement measures}
Using the reconstructed density matrix, we are able to calculate a number of metrics describing the entangled state.
Values ``unconditioned'' and ``conditioned'' on success probability are shown in Table~\ref{tab:entmeas}.
Some of these are useful for comparison with other experiments or when considering, for example, choices among purification schemes.
Entanglement fidelity, concurrence, and purity have been calculated according to standard definitions \cite{Plenio2006}.
The entanglement generation rate $R$ is the rate of experimental repetition, appending additional time required to avoid failed runs (which could in principle be heralded against) in the conditioned case, and inclusive of all experimental runs in the unconditioned case.
As detailed in the main text, this value is limited by the re-initialization time of the system.
The generation rate of maximally entangled qubit pairs, the ebit rate, represents the rate at which an optimal purification scheme \cite{Horodecki2009} could generate a maximally entangled pair from many pairs identical to ours, and can be calculated using the logarithmic negativity \cite{Plenio2005}, $E_\mathcal{N}$:
\begin{equation}
    R_e = p_\text{s,ent} \cdot R \cdot E_\mathcal{N}.
\end{equation}
Success probability $p_\text{s,ent}$ enters in both the logarithmic negativity (where lower $E_\mathcal{N}$ demands more pairs in the unconditioned case) and the generation rate (where overall generation of higher-fidelity entanglement is slower).
\begin{table*}[tb]
    \centering
    \setlength{\tabcolsep}{8pt}
    \begin{tabular}{l c c c} 
    \hline\hline
    Value &  & Conditioned (measured) & Unconditioned (estimated) \\
    \hline
    Success probability                   & $p_\text{s,ent}$            & $0.78\pm0.04$     & $1$ \\
    Fidelity                              & $\mathcal{F}_\mathrm{Bell}$ & $0.77\pm0.02$     & $0.61\pm0.04$ \\
    Concurrence                           & $\mathcal{C}$               & $0.66\pm0.03$     & $0.51\pm0.04$ \\
    Purity                                & $\gamma$                    & $0.66\pm0.03$     & $0.40\pm0.04$ \\
    Generation rate $(\upmu\text{s}^{-1})$& $R$                         & $1/(140 \pm 10)$  & $1/(110 \pm 10)$ \\
    Logarithmic negativity                & $E_\mathcal{N}$             & $0.66\pm0.03$     & $0.30\pm0.04$ \\
    ebit rate $(\text{kebit}/\text{s})$   & $R_e$                       & $4.7 \pm 0.5$     & $2.7 \pm 0.4$ \\
    \hline\hline
    \end{tabular}

    \caption{
    \label{tab:entmeas}
    {\bf Measures of entanglement.}
    All comparison values are with respect the maximally-entangled Bell state, $\left( \ket{10}+\ket{01} \right ) / \sqrt{2}$. ``Unconditioned'' values are those metrics taken for the estimated unconditioned density matrix reconstructed according to the process in Section \ref{sec:ent_succ_cond}, \emph{Success rate conditioning}.
    }
\end{table*}

\subsection{Cavity evolution due to the Kerr effect}
The single-photon Kerr effect in one cavity \cite{Haroche2006,Kirchmair2013} is described by the Hamiltonian
\begin{equation}
    H_\text{Kerr} = \frac{\chiaa}{2} \modea^{\dagger 2} \modea^2.
\end{equation}
This nonlinearity causes a unitary evolution in which each photon number acquires phase at a rate $n\chiaa$. On long time scales the state undergoes cyclic evolution at revival times $2\pi/\chiaa$.
On shorter times, the evolution leads to `smearing' of the phase information, with a phase collapse time that depends on the mean number of photons, $\bar{n}$, in the cavity, $T_{\text{collapse}} = \pi/(2 \sqrt{\bar{n}} \chiaa)$.
In our system, $\chiaa = 2\pi \times 8.8$ $(5.3)\,\text{kHz}$ for sender (receiver), and thus $T_{\text{collapse}} \approx \bar{n}^{-1/2} \times 28$ $(47)\,\upmu\text{s}$. As the collapse times are significantly longer than the transfer time, the effect of the Kerr evolution is a relatively weak distortion of the state.

\paragraph*{Correction in fidelity estimation}
The Kerr effect manifests itself for received states with photon number $\bar{n} > 1$, such as those of the binomial code (\figref4A).
It is only apparent in the Wigner function for states that are not radially symmetric (and contain more than one photon number).
Because the Kerr evolution is deterministic, it does not in and of itself lead to a loss of information, so long as its value is known.
In order to produce a fidelity representative of the effect of the transfer protocol on the prepared states, and not merely the effect of Kerr at one end or the other, we compare the reconstructed density matrix of the prepared state with a reconstructed density matrix of the received state to which an in-software correction has been applied (\figref\ref{fig:binom_expanded}).
This operation is simply the Kerr unitary associated with $H_\text{Kerr}$ above, with an effective Kerr $\chiaa^\text{eff}$, equal to approximately the average of $\chiaa$ in each system.
In order to determine the nominal value of $\chiaa^\text{eff}$ to use for the software correction unitary, we perform a global minimization of the average fidelity over all six states prepared in the manifold with respect to a single Kerr evolution applied to each state. 
We find $\chiaa^\text{eff}=2\pi \times 10.8\,\text{kHz}$ for the transfer time of $6\,\upmu\text{s}$.

Since Kerr evolution does not commute with photon loss, the uncertainty in exactly when the loss occurred does lead to a small unrecoverable loss of information, which manifests as dephasing of the cavity state. 
Because the Kerr is fairly small and the transfer time quite short, this produces a small effect on the measured average fidelity, around 1--2\%. 

\paragraph*{Correction in future devices}
The technique applied in software here can also be applied experimentally with a unitary, and can be combined with the operations necessary to perform error correction.
Though our system parameters inhibit this operation without introducing unnecessary error, a Kerr-reversal unitary could be applied to the receiving cavity using the nominal value of average Kerr, $\chiaa^\text{eff}$.
Undesired Kerr evolution can be suppressed in future experiments by reducing the magnitude of the Kerr effect and by increasing the bandwidth of the conversion process.

\subsection{Model of photon loss}
The model assumes a pure photon loss channel $\Psi$ characterized by the single-photon efficiency $\eta$.
Our measurements produce $\eta = 0.74 \pm 0.03$, which we incorporate in the model.

\paragraph*{Simulation}
We apply the following model of photon loss to ideal or prepared states or density matrices in order to generate density matrices expected in the presence of $\Psi$.
This takes the form of a beamsplitter operation, in which the initial state interferes with an ancillary system in the vacuum state at a variable beamsplitter incidence angle $\theta$.
First, we convert the loss probability $p_\text{loss} = 1 - \eta$ into the interference angle $\theta$ according to
\begin{equation}
\theta = 2 \arccos{\sqrt{1-p_\text{loss}}}.
\end{equation}
Then, the beamsplitter Hamiltonian (\eqnref\ref{eq:bs_theta}) is applied to the product of the states, entangling the Hilbert space of interest with some environmental degree of freedom.
We then trace over the ancillary system to obtain a lossy mixed state.

\paragraph*{Other sources of loss}
While our model of photon loss agrees significantly with the measured process (Section \ref{sec:choi}), other error mechanisms may explain the remaining uncharacterized infidelity.
We have investigated two such mechanisms: decoherence and thermalization.

For simplicity, we chose to apply our models for decoherence and thermalization to the single-photon encoding.
To test decoherence, we construct a positive-operator valued measure (POVM) that includes the action of both photon loss and decoherence.
In this model we represent decoherence using the Pauli operator $\sigma_z$ and photon loss using $\sigma_+ = \sigma_x + i \sigma_y$.
We minimize the average infidelity between the measured states and states resulting from this model, over the probability space of these two mechanisms.
This routine returns a minimum corresponding with a small amount of decoherence that changes the average fidelity (and the fidelity of any single state) by $<1\%$.
Therefore, this effect alone does not account for the $\sim 5\%$ unexplained by the non-unity process fidelity.

The potential effect of thermalization or ``photon gain'' (equilibrating, with some probability, to a thermal state at an unspecified point during the transfer process) is evaluated using a similar comparison.
Much like the model of photon loss imposes a beamsplitter on the transmitted state with vacuum at the opposite input port, the thermalization process is modeled using a thermal state as the joining state.
Including photon gain in the loss model would likely require a reduction in the assumed efficiency due to photon loss, since the same measured states must be ultimately matched.
The model we construct allows for a uniform photon number of the thermal bath, $\bar{n}$, as well as some weight with which the process acts.
Minimizing under this model to best match the measured states, we find an improvement in mean fidelity $<1\%$.
Again, the effect of photon gain alone cannot account for the infidelity of the transfer to a pure photon loss process.

\paragraph*{Octahedron loss trajectories}
The Bloch sphere representations of Fock and binomial encodings are shown in \figref2 and 3 of the main text.
There, we have used this model to produce a continuous ``trajectory'' of points on the Bloch sphere corresponding with the expected states given a prepared state subjected to an transfer protocol with variable loss ($\Psi(\eta)$).
Each trajectory represents the path that the prepared state would take if it were allowed to decay according to our photon loss model.

\begin{figure}[htp]
    \center
    \includegraphics{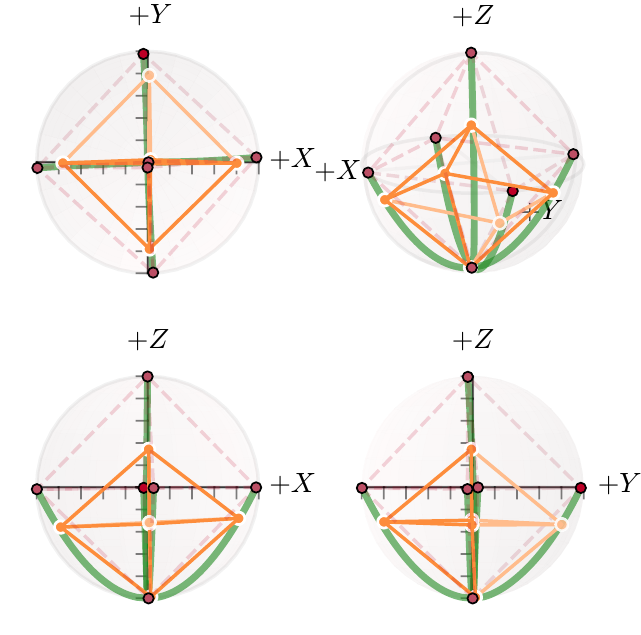}
    \caption{ \label{fig:bloch_fock_complete}
    {\bf Fock encoding loss trajectories.}
    The prepared (red) and measured (orange) points correspond with six cardinal states in the single-photon encoding, and are connected to form an octahedron representative of the manifold.
    The green lines extending from each prepared-state point towards $\ket{0}$ are loss trajectories for varying transfer efficiency from $\eta=0$ to $\eta=1$.
    A three-dimensional perspective corresponding with \figref3B (upper right) as well as third-angle projections (centered on the $XZ$ plane) contain the same data.
    }
\end{figure}

\begin{figure}[htp]
    \center
    \includegraphics{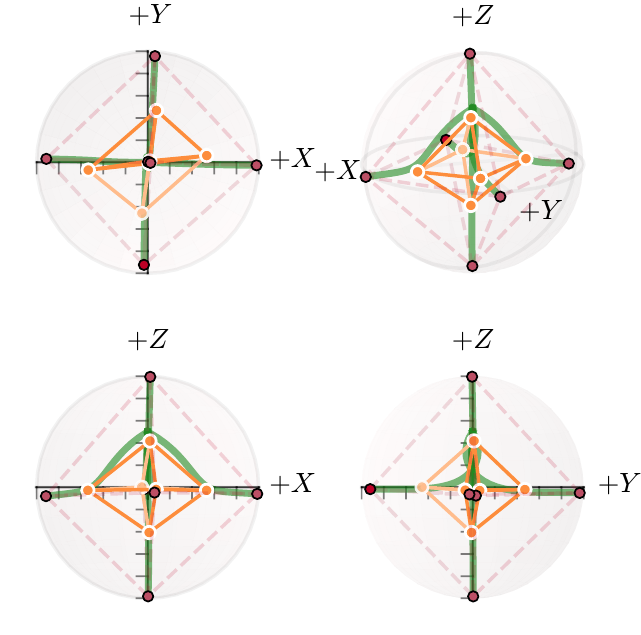}
    \caption{ \label{fig:bloch_binomial_complete}
    {\bf Binomial encoding loss trajectories.}
    The prepared (red) and measured (orange) points correspond with six cardinal states in the binomial encoding, and are connected to form an octahedron representative of the manifold.
    The green lines extending from each prepared-state point towards $(0,0,0.5)$ are loss trajectories for varying transfer efficiency from $\eta=0$ to $\eta=1$.
    A three-dimensional perspective corresponding with \figref4B (upper right) as well as third-angle projections (centered on the $XZ$ plane) contain the same data.
    }
\end{figure}

In the experiment, this loss potentially occurs throughout the transfer process, distributed spatially and temporally.
Values of $p_\text{loss}$ up to the measured efficiency are included there; the full span of $p_\text{loss} \in (0,1)$ are shown in \figref\ref{fig:bloch_fock_complete} and \figref\ref{fig:bloch_binomial_complete}.
In the Fock encoding, photon loss has the effect of reducing all cardinal states to $\ket{0}$ monotonically.
In the binomial encoding, photon loss reduces cardinal points towards the center of the Bloch sphere uniformly, because single-photon errors cause of portion of the state to leave the even-parity codespace.

For larger $\eta$, the cardinal points begin to turn towards the point $(0,0,0.5)$ on the Bloch sphere, since the $\ket{1}_{L} = \left(\ket{0}+\ket{4}\right)/\sqrt{2}$ state overlaps with the vacuum state and $\ket{0}_{L} = \ket{2}$ does not.
Note that our received states appear in the uniform-shrinkage regime, which corresponds to loss small enough that single photon loss errors dominate over higher-order effects.

\paragraph*{In-software error correction}
The green line in \figref4C is calculated by applying the loss model described to the six cardinal states of the binomial encoding. 
The fidelity of the resulting (partially mixed) density matrices are compared to that of each original ideal state, and the average infidelity is reported. 
The gold line in \figref4C is then calculated from these density matrices by applying an error-correction procedure \cite{Michael2016}.

The error-correction procedure is implemented as follows: the density matrix for a given state is split into two components: the even and odd photon number parity subspaces.
A different correction unitary is applied to each subspace.
To correct for photon loss, the odd-parity subspace is corrected via the operation:
\begin{equation}
|1\rangle \rightarrow |2\rangle, \qquad |3\rangle \rightarrow \frac{1}{\sqrt{2}}\left(|0\rangle + |4\rangle\right)
\end{equation}
The correction in the even-parity subspace takes the form of a rotation that adjusts the relative weights of the $|0\rangle$ and $|4\rangle$ states:
\begin{equation}
\begin{aligned}
&|2\rangle \rightarrow |2\rangle,\\
&|0\rangle \rightarrow \cos\theta_\mathrm{c}|0\rangle + \sin\theta_\mathrm{c}|4\rangle,\\
&|4\rangle \rightarrow -\sin\theta_\mathrm{c}|0\rangle + \cos\theta_\mathrm{c}|4\rangle\\
\end{aligned}
\end{equation}
This is to account for the fact that the no-parity-jump event alters the relative probability amplitude of the state $\left(|0\rangle + |4\rangle\right)/\sqrt{2}$.
The two components of the density matrix are then recombined with their respective probabilities unchanged, resulting in a higher-purity corrected density matrix.
The optimum rotation angle $\theta_\mathrm{c}$ depends on the probability of loss.
For each value of the transfer inefficiency, we calculate the angle of rotation that minimizes the average infidelity over the manifold.
This average infidelity is reported in \figref4C.
This procedure is identical to the experimental implementation that will be needed to perform this correction in hardware: measure parity; apply one of two correction unitaries, conditioned on the result of the measurement; and perform unconditional tomography.

\paragraph*{Modifications to allow error correction}
Error-correction of the transfer cannot be performed effectively in the current experimental sample.
Because a single transmon simultaneously provides the functions of state preparation and cavity tomography as well as supplying the nonlinearity needed for the conversion process, it is not possible to optimize the hardware parameters for all of these essential features.
Most notably, a large conversion rate between each memory and output mode requires a large cross-Kerr $\chiab$.
However, this means that resonantly driving the output mode into a coherent state (as is done during transmon readout) dephases the cavity due to the dispersive shift combined with uncertainty in the occupation of the output mode \cite{Gambetta2006}.
Therefore, parity measurement without significant dephasing of the cavity is impossible in the current configuration.
As suggested in the main text, the addition of a dedicated, separate chip with ancilla transmon and readout mode, also with small cross-Kerr between readout and cavity, would allow for high-fidelity QND parity measurement \cite{Ofek2016}, while maintaining the ability to rapidly convert between memory and propagating modes.
This modular distribution of functionality may also allow modifications to the conversion side of the device to mitigate some of the non-idealities discussed above; for example, some nonlinear element other than a transmon can be used to facilitate conversion \cite{Flurin2015,Frattini2017}.

\section{Supplemental data}

\subsection{Transfer of states in the single-photon Fock basis}
\begin{figure*}[tpb]
    \center
    \includegraphics{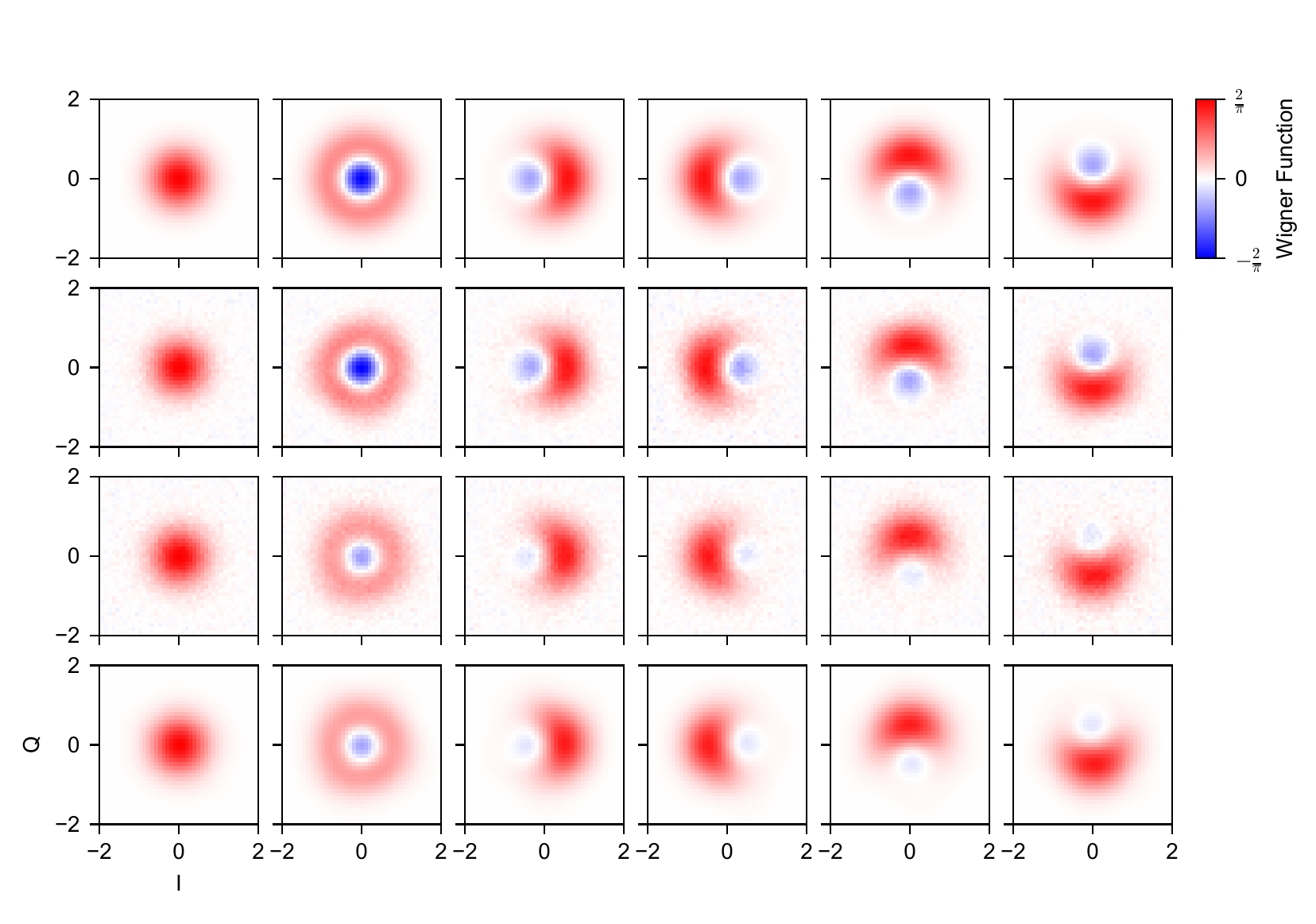}
    \caption{\label{fig:fock_expanded}
    {\bf Full Fock manifold Wigner functions.}
    Six cardinal states in the Fock encoding.
    Left to right: $|{-}Z\rangle$, $|{+}Z\rangle$, $|{+}X\rangle$, $|{-}X\rangle$, $|{+}Y\rangle$, $|{-}Y\rangle$.
    Top to bottom: ideal, measured prepared, measured received, reconstructed received.
    All fidelities are computed using reconstructed density matrices.
    }
\end{figure*}

See \figref\ref{fig:fock_expanded} for measured and reconstructed Wigner functions for all states prepared in the Fock encoding.

\subsection{Transfer of states in a bosonic code}
\begin{figure*}[tpb]
    \center
    \includegraphics{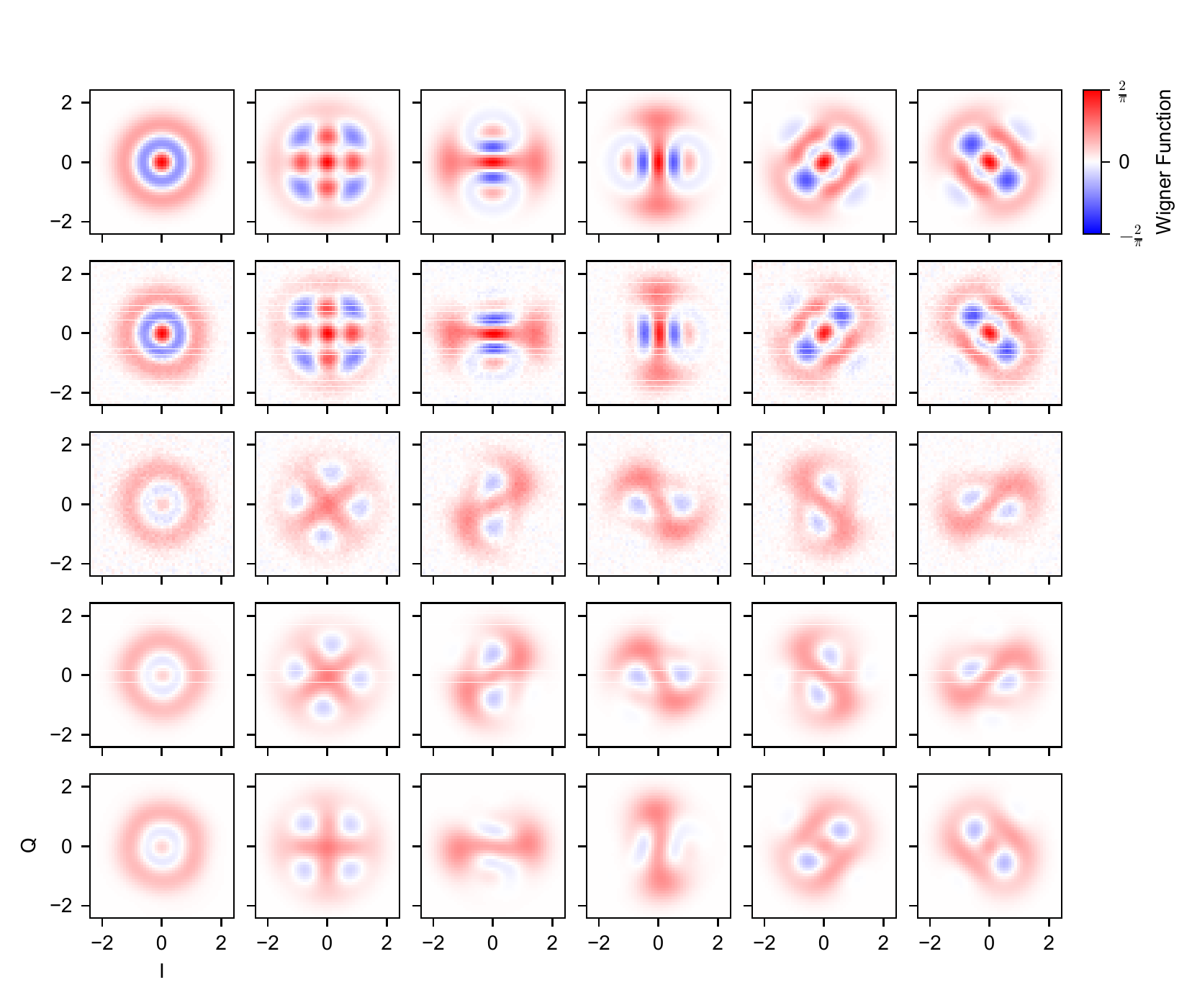}
    \caption{ \label{fig:binom_expanded}
    {\bf Full Binomial manifold Wigner functions.}
    Six cardinal states in the Binomial encoding.
    Left to right: $|{-}Z\rangle$, $|{+}Z\rangle$, $|{+}X\rangle$, $|{-}X\rangle$, $|{+}Y\rangle$, $|{-}Y\rangle$.
    Top to bottom: ideal, measured prepared, measured received, reconstructed received, reconstructed received with Kerr correction.
    All fidelities are computed using reconstructed density matrices.
    }
\end{figure*}

See \figref\ref{fig:binom_expanded} for measured and reconstructed Wigner functions for all states prepared in the Binomial encoding.

\subsection{Process matrix}
\label{sec:choi}
\begin{figure*}[tpb]
    \center
    \includegraphics{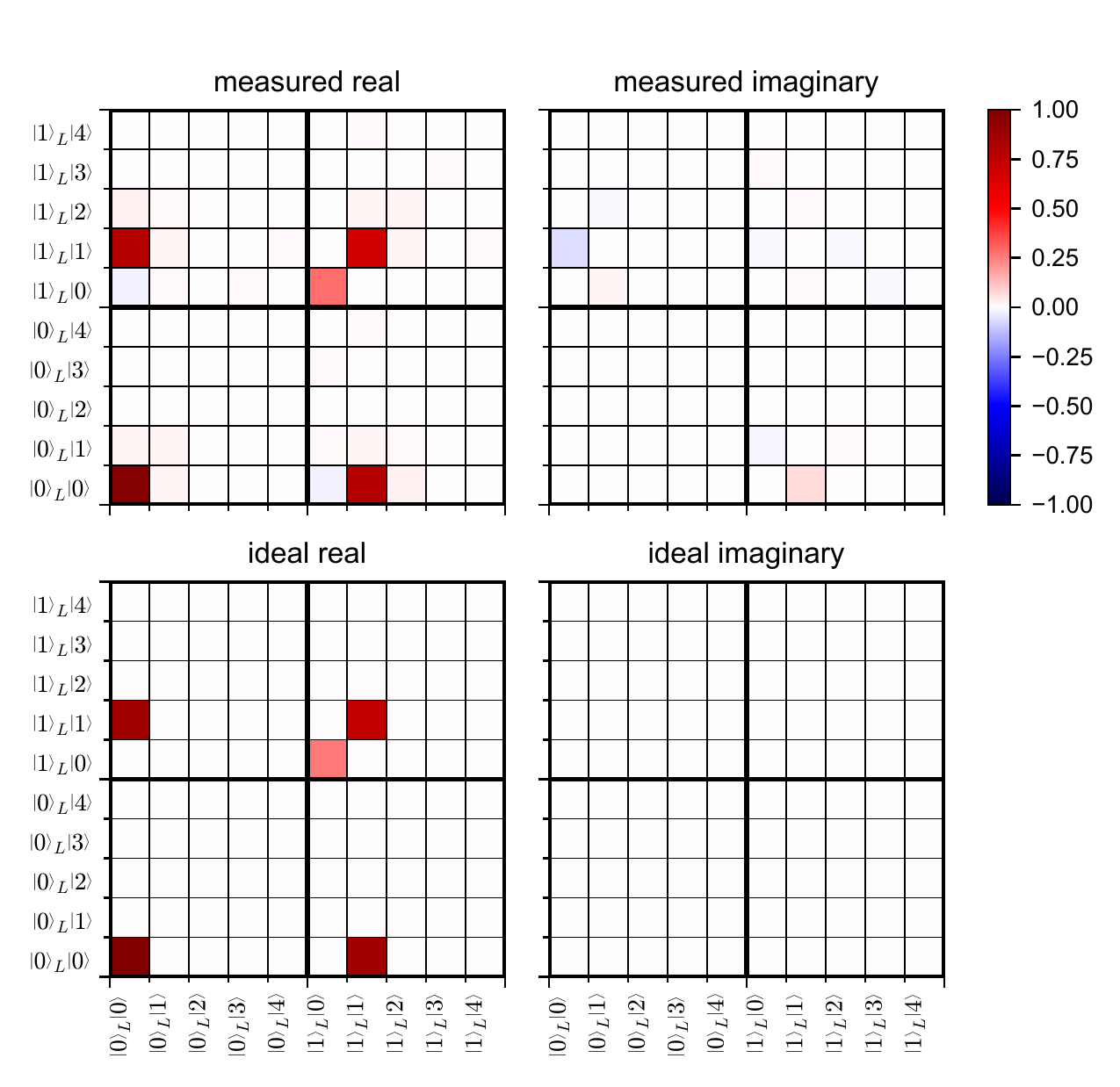}
    \caption{\label{fig:choi}
    {\bf Process matrix.}
    {\bf Top:} Reconstructed process matrix for state transfer in the Fock encoding.
    The subscript $L$ denotes the logical qubit space, while the kets with no subscript reside in the physical space of the cavity.
    {\bf Bottom:} Ideal process matrix assuming transmission loss of the measured $\eta=0.74$.
    }
\end{figure*}

The process matrix $\chi$ defines the mapping from the ideal state density matrix $\rho_i$ to the final, received density matrix $\rho_f$.
In this definition, $\rho_i$ is a two-dimensional logical qubit state in the manifold $\{|0\rangle_L, |1\rangle_L \}$, while $\rho_f$ is in the high-dimensional physical Hilbert space of the cavity: $\{|0\rangle, |1\rangle,... |d-1\rangle\}$.
For our purposes, we truncate this space at dimension $d=5$; for the Fock encoding, no significant population is found outside this space upon reconstruction.
The process matrix is then defined as the map from $2\times2$ logical density matrices to $d\times d$ physical density matrices.
The reconstructed process matrix $\chi_m$ is visualized in \figref\ref{fig:choi} and compared to the ideal process $\chi_i$ assuming only the measured loss $\eta=0.74$.
The process fidelity $\mathcal{F}_\mathrm{process}=\frac{1}{4}\text{Tr}(\sqrt{\sqrt{\chi_i} \chi_m \sqrt{\chi_i}})^2 =0.95$, implying very close agreement with our model of photon loss alone.
We believe the source of this $0.05$ infidelity also produces the $\sim 0.05$ average state infidelity (relative to what is expected from the measured inefficiency; as stated in the main text, $\mathcal{F}_\mathrm{avg} = 0.87 \pm 0.04$, while we expect $0.91$.)

\end{document}